\documentclass{emulateapj}

\usepackage{graphicx}
\usepackage{epsfig}
\usepackage{natbib}
\usepackage{multirow}
\usepackage{amsmath,amssymb}
\usepackage{rotating}
\usepackage{hyperref}

\begin{document}
\submitted{ApJ in press}
\journalinfo{Draft version \today}

\title{Faint Submillimeter Galaxy Counts at 450 microns}
\author{Chian-Chou Chen\altaffilmark{1}, Lennox L. Cowie\altaffilmark{1}, Amy J. Barger\altaffilmark{1,2,3},  Caitlin. M. Casey\altaffilmark{1,$\star$}, Nicholas Lee\altaffilmark{1}, David B. Sanders\altaffilmark{1}, Wei-Hao Wang\altaffilmark{4}, Jonathan P. Williams\altaffilmark{1}}
\altaffiltext{1}{Institute for Astronomy, University of Hawaii, 2680 Woodlawn Drive, Honolulu, HI 96822.}
\altaffiltext{2}{Department of Astronomy, University of Wisconsin-Madison, 475 North Charter Street, Madison, WI 53706.}
\altaffiltext{3}{Department of Physics and Astronomy, University of Hawaii, 2505 Correa Road, Honolulu, HI 96822.}
\altaffiltext{4}{Academia Sinica Institute of Astronomy and Astrophysics, P.O. Box 23-141, Taipei 10617, Taiwan.}
\altaffiltext{$\star$}{Hubble Fellow}
\subjectheadings{cosmology: observations|  galaxies: evolution  |  galaxies: formation  |  submillimeter}

\begin{abstract}
We present the results of SCUBA-2 observations at 450\,$\mu$m and 850\,$\mu$m of the field lensed by the massive cluster A370. With a total survey area $>$ 100 arcmin$^2$ and 1\,$\sigma$ sensitivities of 3.92 and 0.82 mJy/beam at 450 and 850\,$\mu$m respectively, we find a secure sample of 20 sources at 450\,$\mu$m and 26 sources at 850\,$\mu$m with a signal-to-noise ratio $>$ 4. Using the latest lensing model of A370 and Monte Carlo simulations, we derive the number counts at both wavelengths. The 450\,$\mu$m number counts probe a factor of four deeper than the counts recently obtained from the {\it Herschel Space Telescope\/} at similar wavelengths, and we estimate that $\sim$47--61\% of the 450\,$\mu$m extragalactic background light (EBL) resolved into individual sources with 450\,$\mu$m fluxes greater than 4.5\,mJy. The faint 450\,$\mu$m sources in the 4\,$\sigma$ sample have positional accuracies of 3 arcseconds, while brighter sources (signal-to-noise $>6\,\sigma$) are good to 1.4 arcseconds. Using a deep radio map ($1\,\sigma\sim 6\,\mu$Jy) we find that the percentage of submillimeter sources having secure radio counterparts is 85\% for 450\,$\mu$m sources with intrinsic fluxes $>6$\,mJy and 67\% for 850\,$\mu$m sources with intrinsic fluxes $>4$\,mJy. We also find that 67\% of the $>4\,\sigma$ 450\,$\mu$m sources are detected at 850\,$\mu$m, while the recovery rate at 450\,$\mu$m of $>4\,\sigma$ 850\,$\mu$m sources is 54\%. Combined with the source redshifts estimated using millimetric flux ratios, the recovered rate is consistent with the scenario where both 450\,$\mu$m and 20\,cm emission preferentially select lower redshift dusty sources, while 850\,$\mu$m emission traces a higher fraction of dusty sources at higher redshifts. We identify potential counterparts in various wavelengths from X-ray to mid-infrared and measure the multiwavelength photometry, which we then use to analyze the characteristics of the sources. We find three X-ray counterparts to our robust submillimeter sample (S/N $>$ 5), giving an active galactic nucleus fraction for our 450 (850)\,$\mu$m sample of 3/8 (3/9) or 38\% (33\%). We also find a correlation between the $K_s$-band and the 850\,$\mu$m/20~cm flux ratio. 
\end{abstract}

\section{Introduction}
Based on studies of the integrated light originating from outside the Milky Way galaxy---the extragalactic background light (EBL)---we now know that there is a comparable amount of light absorbed by dust and re-radiated in the far-infrared (FIR)/submillimeter (100\,$\mu$m $< \lambda <$ 1000\,$\mu$m) as there is seen directly in the optical/UV (\citealt{Puget:1996p2082,Fixsen:1998p2076, Dole:2006p9898}), which indicates that much of the star formation and active galactic nucleus (AGN) activity in the Universe is hidden by dust. Understanding dusty star formation is therefore critical to a full understanding of galaxy formation. 

The first step in such work is identifying the individual dusty galaxies. Observations have been made towards this goal using both the small space-based Herschel Space Observatory (hereafter {\it Herschel}; \citealt{Pilbratt:2010lr}) and, at longer wavelengths, ground-based telescopes (e.g., the James Clerk Maxwell Telescope (JCMT)). However, the fraction of sources that can be identified with such observations has a fundamental limit due to the poor resolution. This limit is known as the confusion limit (\citealt{Scheuer:1957fk,Condon:1974qy,Hogg:2001uq}), which is the maximum sensitivity an observation is able to reach for a given beam size due to the overlap of sources on the map. Recent observations carried out by {\it Herschel} at 250, 350, and 500\,$\mu$m can only resolve a small amount of the EBL (15, 10, and 6\%; \citealt{Oliver:2010p11204}) due to the confusion limit. With a 15 meter dish, the JCMT blank-field submillimeter surveys with SCUBA (\citealt{Holland:1999fk}) only resolved $\sim$20\%--30\% of the 850\,$\mu$m EBL into distinct, bright submillimeter galaxies (SMGs) with S$_{850\,\mu m} >$ 2\,mJy (e.g., \citealt{Barger:1998p13566, Hughes:1998p9666, Barger:1999p6485,Eales:1999p9715, Coppin:2006p9123}) before reaching the confusion limit. The most recent surveys using LABOCA (\citealt{Siringo:2009lr}) on the APEX telescope at 870\,$\mu$m obtained similar results (Wei\ss~et al. 2009). 

The most fundamental consequence of the confusion limit is that the poor resolution at 850\,$\mu$m prevents the study of faint SMGs below 2 mJy in blank fields. A small number of fainter SMGs with intrinsic fluxes between 0.1 and 2 mJy have been detected from observations of the fields of massive lensing clusters. In these fields, the intrinsically faint fluxes of background sources are gravitationally amplified to a detectable level. The sources found in this way contribute more than 50\% of the 850\,$\mu$m EBL (\citealt{Smail:1997p6820,Smail:2002p6793, Cowie:2002p2075, Knudsen:2008p3824}). However, the positional uncertainties of the 850\,$\mu$m sources can cause large uncertainties in the amplifications and in the intrinsic source fluxes (\citealt{Chen:2011p11605}). It is therefore essential to conduct surveys on lensing cluster fields with as high a spatial resolution as possible to circumvent the confusion limit and to construct the most accurate possible number counts distribution. 

The poor resolution (e.g., 14$''$ FWHM on the JCMT at 850\,$\mu$m) also makes identifying the counterparts in other wavelengths difficult and time consuming. Deep, high-resolution radio, mid-infrared (MIR), optical, UV, and X-ray observations have been used to identify candidate counterparts to the SMGs and to trace their star formation and active galactic nucleus (AGN) activity (\citealt{Barger:2000p2144, Chapman:2005p5778, Pope:2006p8076, Ivison:2007lr, Georgantopoulos:2011mz}). However, direct high-resolution submillimeter interferometric observations have shown that due to their clustered nature (\citealt{Weis:2009qy}), SMGs are likely to be resolved into multiple sources and that different tracers identify different counterparts based on their own biases (e.q., \citealt{Wang:2011p9293, Barger:2012lp, Smolcic:2012lr, Karim:2012lr}).

Many of these problems can be avoided by observing at 450\,$\mu$m rather than at 850\,$\mu$m. The higher spatial resolution at 450\,$\mu$m means that we can detect several times more sources before hitting the confusion limit, and the more accurate positions allow us to better determine the amplifications when observing lensed sources. In addition, while 850\,$\mu$m observations have the unique advantage of tracing the star-forming galaxies to very high redshifts ($z\sim10$), thanks to the negative $K$-correction (\citealt{Blain:2002p8120}), it has been argued that owing to the location on the Rayleigh-Jeans tail of the blackbody spectral energy distribution (SED), the 850\,$\mu$m selection is biased against high dust temperature sources (\citealt{Blain:2004fj,Chapman:2004kx,Casey:2009p13240}). To mitigate this selection effect, observations at rest-frame wavelengths close to the peak of the blackbody SED ($\sim 100-200~\mu$m) that are less affected by the dust temperature are needed. The SPIRE instrument mounted on {\it Herschel} is designed to do this job; however, its ability to detect sources is again heavily restricted by the confusion limit due to the small aperture size of the telescope.

450\,$\mu$m mapping was very inefficient with SCUBA (\citealt{Chapman:2002p9010, Smail:2002p6793}). Based on a handful of extracted sources ($<$ 10) with 450\,$\mu$m fluxes above 10 mJy, only 15\% of the 450\,$\mu$m EBL was resolved (\citealt{Smail:2002p6793}). The advent of SCUBA-2 (\citealt{Holland:2006fk}) on the JCMT with its two orders of magnitude more pixels/bolometers than on SCUBA greatly boosts our ability to resolve the bulk of the 450\,$\mu$m EBL and to detect galaxies not biased by dust temperature. SCUBA-2 observes simultaneously at 450 and 850\,$\mu$m via a dichroic beamsplitter. Unlike SCUBA, the bolometers on SCUBA-2 are optimized for observing at 450 and 850\,$\mu$m, and the filter transmission, especially at 450\,$\mu$m, is much better than that of SCUBA. Moreover, the field-of-view of SCUBA-2 is more than 10 times larger than that of SCUBA. With all of these new features, SCUBA-2 is an order of magnitude faster than SCUBA, which makes it possible to finally study the 450\,$\mu$m population in detail. The amount of FIR EBL that can be resolved by SCUBA-2 is unprecedented as we show later in the paper, and demonstrates both the power of SCUBA-2 and also the promise of the next generation of submillimeter telescopes (CCAT).

Here we describe the results that we have obtained on the lensing cluster field Abell 370 (A370) from our ongoing SCUBA-2 program to survey a sample of massive galaxy clusters at 450\,$\mu$m. The goals of this program are to resolve the 450\,$\mu$m EBL as fully as possible, to efficiently identify the counterparts of the 850\,$\mu$m sources, and, most important of all, to study the nature of the SMG population in as unbiased a fashion as possible. In Section~2, we discuss our SCUBA-2 observations, data reduction, and ancillary data.  Since the results involve a relatively new instrument we describe the methodology in detail. Readers who are primarily interested in the results can skip directly to section 3. In Section~3, we describe our source extraction and present our catalog. In Section~4, we determine the number counts and the amount of the EBL that our data resolve at both 450 and 850\,$\mu$m.  In Section 5, we discuss the properties of the sources at submillimeter and radio wavelengths. Using the rich set of ancillary data on this field from the radio to the X-ray, in Section~6, we analyze the properties of the counterparts to the submillimeter sources, construct color-color and color-flux diagrams, and examine the nature of both 450 and 850\,$\mu$m sources. Finally, we summarize our results in Section~7. Throughout the paper, we adopt the AB magnitude system for the photometry. We also assume the Wilkinson Microwave Anisotropy Probe cosmology of H$_0$ = 70.5~km~s$^{-1}$~Mpc$^{-1}$, $\Omega_M = 0.27$, and $\Omega_\Lambda = 0.73$ (\citealt{Larson:2011ys}).  

\begin{figure*}
 \begin{center}
    \leavevmode
      \includegraphics[width=1 \textwidth]{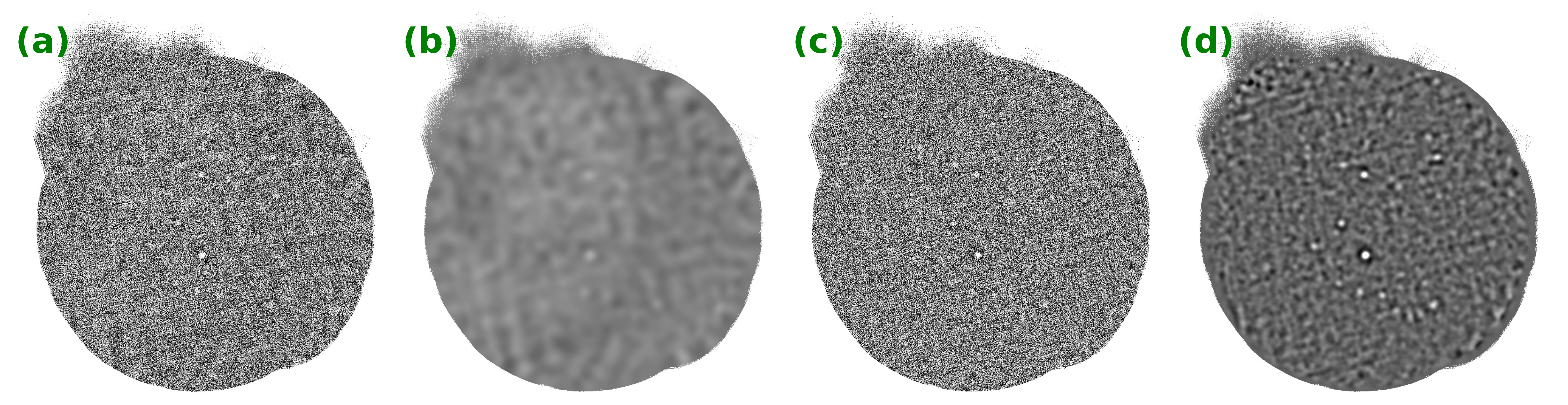}
       \caption{Demonstration 850\,$\mu$m maps for post-processing. (a) Map created from DIMM. (b) Map in (a) convolved with a broad Gaussian (FWHM = 30$''$ for 850\,$\mu$m). (c) Map in (a) subtracted from map in (b). (d) Result of performing matched-filter on map in (c). Note that these are all S/N maps scaled to maximize the visibility of the sources.}
     \label{a370_mtchfltrprcs}
  \end{center}
\end{figure*}

\section{Observations, Data Reduction and Ancillary Data}

We carried out the SCUBA-2 observations in December 2011 and in February and March 2012 (Program IDs: M11BH26A and M12AH26A) under the best weather conditions (band 1, tau$_{\rm 225~GHz}<0.05$). A370 has a large Einstein ring with an angular diameter size of 3$'$. To exploit the strong magnification and to efficiently use any available time to discover faint sources, we adopted the CV Daisy pattern\footnote[5]{A detailed description of the CV Daisy pattern can be found at \url{http://www.jach.hawaii.edu/JCMT/continuum/scuba2/scuba2\_obsmodes.html\#scan}.} to perform the observations. This observing mode is optimal for an area less than 4$'$. Each scan/cycle takes around 20 minutes to complete. Flatfields were taken at the beginning and end of each scan, however only the first one is used for data reduction. The extinction correction coefficients are obtained from the JCMT water-vapour radiometer (WVM) using the 183\,GHz water line to calculate the precipitable water vapour (PWV) along the line-of-sight of the telescope (\citealt{Dempsey:2012yq}).

We reduced the data using the Dynamic Iterative Map-Maker (DIMM) in the SMURF package from the STARLINK software developed by the Joint Astronomy Centre (\citealt{Jenness:2011lr}) released before July 1, 2012. The program automatically performs dark subtraction, flatfielding, extinction corrections, and data flagging. It also removes the DC steps from each bolometer, performs frequency domain filtering, and concatenates the reduced data from each chip to produce the final sky map. DIMM provides considerable freedom to the user to fine-tune the parameters that control the jobs mentioned above, and those parameters are stored in a configuration file that is included in DIMM during the data reduction. 

There are four detector chips in each waveband with 1280 (40$\times$32) bolometers per chip. We ran DIMM on the individual chips in each scan and used the MOSAIC\_JCMT\_IMAGES recipe in PICARD (Pipeline for Combining and Analyzing Reduced Data; \citealt{Jenness:2008fk}) to mosaic and coadd the maps. We found this method produces better signal-to-noise (S/N) maps than running DIMM on all the data at once. The noise maps and the exposure time maps were also created from DIMM. The noise map is obtained by computing the variance of the data that lands in each pixel. We tested the robustness of the noise map created from DIMM by checking the standard deviation of the S/N maps. If the noise map is robust, then the standard deviation of the S/N map should be close to 1. All the noise maps are accurate with an underestimation less than 5\%. We suspect that the cause of the small underestimation is the correlated noise from the large scale structure caused by the atmospheric noise (see Figure \ref{a370_mtchfltrprcs}(a)), which is ultimately subtracted out in our post-processing technique (Section~2.1).

\subsection{Flux Calibration}
\begin{table*}
\begin{center}
\caption{FWHM of the PSFs and FCFs of the Calibrators}
\scalebox{1}{
\begin{tabular}{ccccc}
 \hline
\hline
Date  &  Calibrators  &  FWHM$_X$   &  FWHM$_Y$   &  FCF\_[BEAMMATCH]   \\
&&(arcsec; 450/850)&(arcsec; 450/850)&(Jy/Beam/pW; 450/850) \\
 \hline
                122611  &  CRL618       &  8.96 / 13.26        &  8.64 / 13.03        &  831.513 / 606.928            \\
                122811  &  Uranus     &  8.74 / 13.06        &  8.41 / 13.23        &  750.590 / 567.313            \\
                        &  Uranus     &  9.41 / 13.40        &  9.51 / 13.94        &  819.377 / 602.322            \\
                020412  &  Uranus       &  8.32 / 12.77        &  9.44 / 13.65        &  697.301 / 568.398            \\
                020512  &  Uranus       &  8.47 / 13.06        &  8.71 / 13.45        &  710.364 / 563.341            \\
\hline
               Average  &               &  8.78 / 13.11        &  8.94 / 13.46        &  761.829 / 581.660            \\
                Stddev  &               &  0.43 / 0.24         &  0.50 / 0.36         &  61.454 / 21.111              \\
\hline
\end{tabular}
}
\label{cal}
\end{center}
\end{table*}

We adopted the configuration file {\it dimmconfig\_bright\_compact.lis} provided by SMURF for DIMM to produce the calibrators' maps in units of picowatts (pW), and we used the CALC\_SCUBA2\_FCF recipe from PICARD to calculate the Flux Conversion Factors (FCFs) for the SCUBA-2 maps. 

The recipe performs a Gaussian fit for the calibrator and computes the total flux in units of pW~arcsecond$^2$. It then divides the expected flux in Jy by the computed flux to obtain the FCF\_[ARCSEC] in units of Jy/pW/arcsecond$^2$. It has been reported that FCF\_[ARCSEC] is very stable from extensive examination of the calibrators observed over the period of SCUBA-2 commissioning and science verification (\citealt{Dempsey:2012yq}). We doubled checked this on our own calibrators and confirmed that the number is indeed stable. 

The majority of SMGs are found to be located at high redshifts ($z>1$), with the median $z\sim2.5$ (\citealt{Chapman:2005p5778, Wardlow:2011qy}). Based on high-resolution interferometric observations, a physical resolution less than a few kpc is needed to resolve individual SMGs (\citealt{Knudsen:2010p9150}). With our 7$\farcs$5 resolution at 450\,$\mu$m, we can barely resolve sources at $z\sim0.1$. Thus, it is safe to assume that SMGs are mostly unresolved point sources at the resolution of our maps. 

To optimize the point source extraction, we adopted the matched-filter algorithm, which is a maximum likelihood estimator of the source strength, for post-processing in order to increase the S/N for point source detection (e.g., \citealt{Serjeant:2003lr}). Assuming S(i,j) and $\sigma$(i,j) are the signal and r.m.s noise maps produced by DIMM and PSF(i,j) is the signal point spread function, the filtered signal map F(i,j) would be 
\begin{equation}
F(i,j) = \frac{\sum_{i,j} [S(i,j) / \sigma(i,j)^2 \times PSF(i,j)]}{\sum_{i,j} [1 / \sigma(i,j)^2 \times PSF(i,j)^2]}, 
\end{equation} 
and the filtered noise map N(i,j) would be 
\begin{equation}
N(i,j) = \frac{1}{\sqrt{\sum_{i,j} [1 / \sigma(i,j)^2 \times PSF(i,j)^2]}}. 
\end{equation} 

Ideally, the PSF for the matched-filter algorithm is a Gaussian normalized to a peak of unity with FWHM equal to the JCMT beamsize at a given wavelength (i.e., $7\farcs5$ at 450\,$\mu$m and $14''$ at 850\,$\mu$m). However, the map produced from DIMM usually has low spatial frequency structures that need to be subtracted off before performing the source extraction. Thus, before running matched-filter, we convolved the map with a broad Gaussian (22$''$ at both 450 and 850\,$\mu$m) normalized to a sum of unity and subtracted this convolved map from the original map. Note that we experimented with different FWHM sizes of the broad Gaussian used to convolve the maps for background subtraction, and we found that the source fluxes and the S/N are not sensitive to the size of the FWHM (with variations lower than 5\%) for reasonable choices. We adopted 22$''$, which gives a  good suppression of the negative signals. In Figure \ref{a370_mtchfltrprcs}, we show the before and after versions of the background subtraction and matched-filter. To determine the fluxes, we processed the PSF used for matched-filter similarly. It becomes a Gaussian with a convolved broader Gaussian subtracted off, which gives a Mexican hat-like wavelet. We adopted the PICARD recipe SCUBA2\_MATCHED\_FILTER for the tasks described above. 

Because of the atmospheric fluctuation, especially at 450\,$\mu$m, the signal distribution is not always a perfect Gaussian with the expected FWHM, which could lead to incorrect flux measurements if we adopt the standard PSF for the matched-filter. However, this uncertainty can be taken into account by going through the same source extraction process on the calibrators as was done on the science maps. We therefore applied matched-filter to the calibrators with the same PSF function used on the science maps, including the background subtraction, and divided the calibrators' expected flux by the after-matched-filter fitted peak flux to obtain FCF\_[BEAMMATCH] in units of Jy/beam/pW. This is the FCF that we used to calibrate our science maps.  

There are several calibrators available each night, and the variation of FCF\_[BEAMMATCH] between calibrators is usually larger than the statistical error. We chose the primary calibrator Uranus most of the time, because of its high S/N. Moreover, compared to the other primary calibrator Mars, the angular size of Uranus is more stable and smaller than the beam size (unresolved), which is ideal for flux calibration for point sources. However, in cases where the phase of Uranus is highly distorted, we considered the secondary calibrator. Table~\ref{cal} shows the 2-D gaussian fitted FWHM on the $x$ and $y$ axes, along with the FCF\_[BEAMMATCH] of all our calibrators. The FWHM of the PSFs and the FCFs are very stable for all our observations (less than 10\% variations), which is evidence that the PSFs at both wavelengths are well determined.

\subsection{Science Maps}
\begin{table}
\begin{center}
\caption{SCUBA-2 observations on A370 field}
\begin{tabular}{lr}
\hline
\hline
 R.A. (J2000; h:m:s)                        &   02:39:53  \\
 DEC. (J2000; $^\circ$:$'$:$''$)                        &  -01:34:38  \\
 Area (850\,$\mu$m; arcmin$^2$)$^{a}$                  &    124.579  \\
 Area (450\,$\mu$m; arcmin$^2$)$^{a}$                  &    121.095  \\
 Central Noise (850\,$\mu$m; 1\,$\sigma$; mJy/beam)  &       0.82  \\
 Central Noise (450\,$\mu$m; 1\,$\sigma$; mJy/beam)  &       3.92  \\
\hline 
& \\
 \multicolumn{2}{l}{$^{a}$ Total area to 3 times the central noise level} \\ 
\end{tabular}
\label{obs}
\end{center}
\end{table}

We adopted the configuration file {\it dimmconfig\_blank\_field.lis} for our science maps. The biggest difference between {\it dimmconfig\_blank\_field.lis} and {\it dimmconfig\_bright\_compact.lis} used for calibrators is that the latter is less aggressive in identifying bad bolometers and DC steps, in order to ease the misidentification between bad bolometers/DC steps and bright sources. 

For each night of data, we used DIMM to reduce the data on the individual chips in each scan and then coadded them to produce one science map and one calibrator map. We performed the matched-filter algorithm on both maps and applied the FCF\_[BEAMMATCH] obtained from the calibrator to the science map. Finally, we coadded all the science maps that were created on a night-by-night basis to produce the final map for source extraction.

Because we subtracted the original signal map from a self-convolved map during the post-processing to get rid of the large scale structure, negative troughs are generated around strong sources (see Figure~1(d)). Those negative troughs may suppress nearby weak sources. We therefore re-did the post-processing by masking out sources with S/N greater than 4 and iterated this procedure until there were no new sources with S/N greater than 4. We then put the masked sources back into the masked images to generate a hybrid map, which is the final map that we used for our analysis. 

\begin{figure}[t]
 \begin{center}
    \leavevmode
      \includegraphics[scale=0.5]{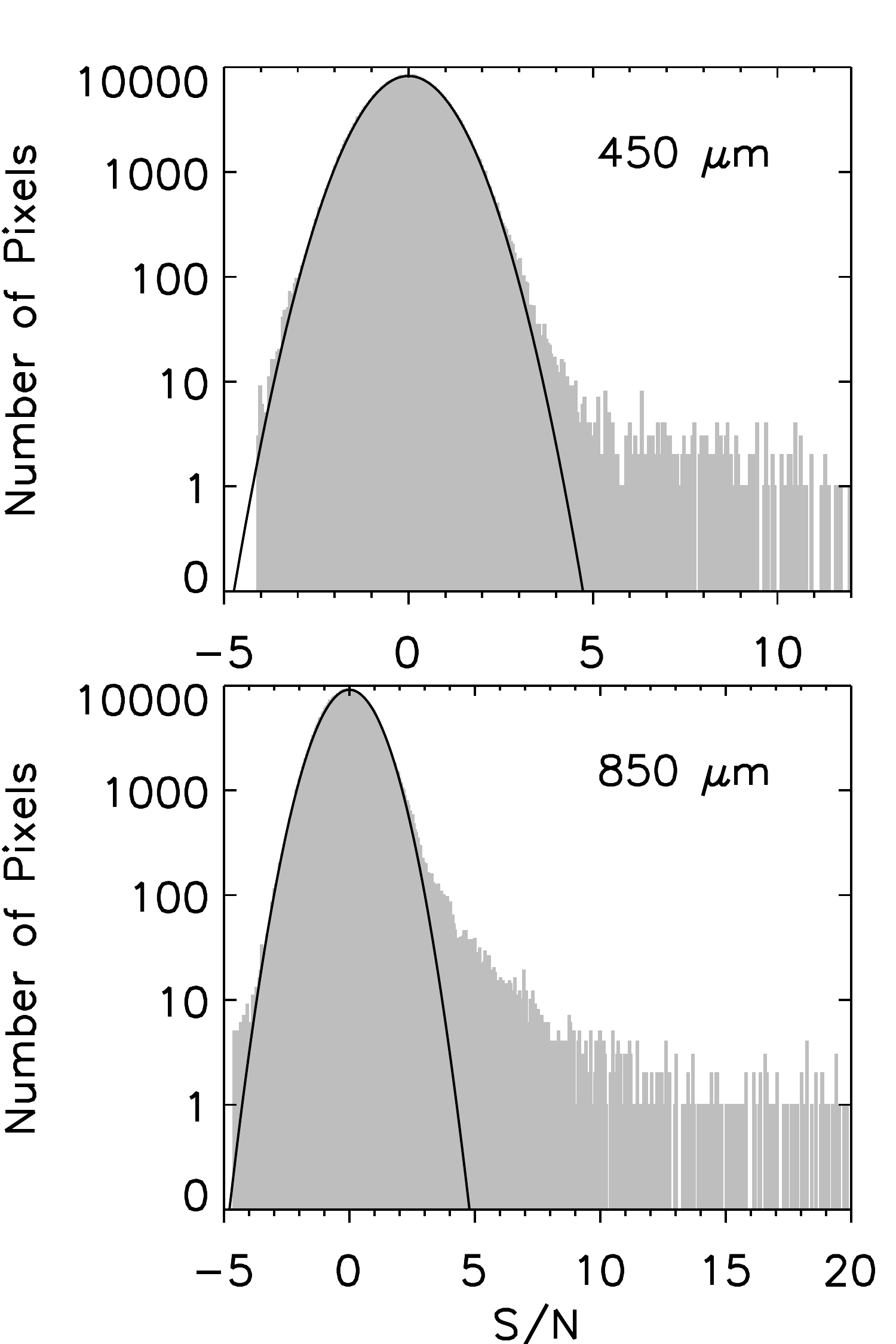}
       \caption{The histograms of the S/N values of the pixels located within the regions where the noise level is less than three times the central r.m.s.\ (gray area). The black curves represent the region of noise. The excess negative signals at 850\,$\mu$m come from one negative source (see discussion in Section~3.3).}
     \label{histo}
  \end{center}
\end{figure}

The noise map created through Equation~2 would inevitably underestimate the noise level, because of the spatially correlated noise (Figure \ref{a370_mtchfltrprcs}(d)). We corrected for this underestimation by multiplying the noise map by the standard deviation of the original S/N map (Equation~1 divided by Equation~2). In Figure~\ref{histo}, we plot the histograms of the final data maps at both wavelengths within the effective regions. The effective regions are the regions where the noise level is less than three times the central r.m.s., meaning the region where the effective exposure time is more than 1/9 of the central exposure time. The black curves represent the region of noise. The excess negative signals at 850\,$\mu$m come from one negative source (see discussion in Section~3.3). We present the final central noise level, along with other observational information on A370, in Table \ref{obs}. 

\subsection{Astrometry Calibration}

SMGs are often traced by 20~cm sources, thanks to the tight and universal FIR-radio flux correlation (\citealt{Helou:1985qy, Condon:1992p6652}). Although this correlation must break down at very high redshifts due to the Compton cooling between the relativistic electrons and the cosmic microwave background photons, so far no clear evidence for evolution of this correlation has been seen to $z\sim 5$  (\citealt{Barger:2012lp}). 

Since most SMGs found so far are located at $z<5$, the radio sources detected in high-resolution maps taken by interferometers are a good sample to use for our absolute astrometry calibration. We searched for the offset of the submillimeter maps  that maximized the stacked submillimeter signal over all the radio sources in the \citet{Wold:2012lr} radio catalog of the field. We found a systematic shift of (0$''$, 2$''$) in our 850\,$\mu$m map and (-1$''$, 1$''$) in our 450\,$\mu$m map. We corrected this small offset before proceeding to the source extraction.

\subsection{Ancillary Data}

A370 is one of the most extensively studied gravitational lensing cluster fields, with its multiwavelength observations from the X-ray to the radio. In Table \ref{mw}, we list all the data on A370 that we used in this paper, along with the corresponding references. 

The {\it Chandra} X-ray survey of the A370 field with 66.6~ks of useful exposure time has a 1\,$\sigma$ sensitivity of $\sim$ 10$^{-15}$~erg~cm$^2$~s$^{-1}$ in the 2--7~keV band and $\sim 3\times$10$^{-16}$~erg~cm$^2$~s$^{-1}$ in the 0.5-2 keV band (\citealt{Bautz:2000qy,Barger:2001uq}). 

The optical $z$-band data were taken by \citet{Hu:2010ys} using Suprime-Cam on Subaru for a study of high-redshift Ly$\alpha$ emitters. The 5\,$\sigma$ sensitivity for a 3$''$ diameter aperture is 25.4 AB magnitude. 

The deep NIR $J, H, K_s$ observations were carried out by \citet{Keenan:2010fr} using ULBCam on the University of Hawaii 2.2~m (UH2.2m) and WFCam on UKIRT. The 5\,$\sigma$ detection limits are 23.5, 23.0, and 23.0 AB magnitude, respectively. We adopted their configuration files and used SExtractor (\citealt{Bertin:1996zr}) to extract sources. 

The A370 field was first surveyed by {\it Spitzer} (\citealt{Werner:2004uq}) in 2006 (PI: Giovanni Fazio; PID: 137) in all the IR Array Camera (IRAC; \citealt{Fazio:2004fj}) bands (3.6-8.0\,$\mu$m).  Data with wider coverage at 3.6 and 4.5\,$\mu$m were later taken in 2009 and 2010 (PI: Eiichi Egami; PID: 60034) when {\it Spitzer} entered the warm phase. We again used SExtractor to extract sources from the {\it Spitzer} maps. We measured the source fluxes using apertures with diameters of 4$\farcs$8 for 3.6\,$\mu$m and 4.5\,$\mu$m and 6$\farcs$0 for 5.8\,$\mu$m and 8.0\,$\mu$m, which are roughly three times the FWHM of the PSFs and are the same as those used in \citet{Wang:2004p2270}. We estimated the sensitivities using Gaussian fits to the fluxes measured at random source-free positions, finding 5\,$\sigma$ limits of 23.3 at 3.6\,$\mu$m and 4.5\,$\mu$m, 23.0 at 5.8\,$\mu$m, and 22.6 at 8.0\,$\mu$m, all in AB magnitudes. 

The deep VLA radio data of the field were taken using both the A and B configurations. The final image reaches a 5\,$\sigma$ limit of 28.5\,$\mu$Jy/beam near the central region with a beam size of $\sim 1\farcs7$ (\citealt{Wold:2012lr}).

\begin{table}
\begin{center}
\caption{A370 Optical, NIR, MIR, FIR, Radio Photometry}
\scalebox{1}{
\begin{tabular}{lll}
 Wavebands           &  Instruments/Telescopes   & References    \\
\hline
X-ray (2--7 keV)	& ACIS-S3/{\it Chandra} 	&	Barger et al. 2001\\
X-ray (0.5--2 keV)	& ACIS-S3/{\it Chandra} 	& 	Barger et al. 2001\\
 $z$                   		& Suprime-Cam/Subaru 	&	Hu et al. 2010 \\
 $J$                   		&  ULBCam/UH2.2m    	&   	Keenan et al. 2010      \\
 $H$                   		&  ULBCam/UH2.2m            & 	Keenan et al. 2010\\
 $K_s$                   	&  WFCAM/UKIRT    		&  	Keenan et al. 2010 \\
 3.6\,$\mu$m          	&  IRAC/{\it Spitzer}   	&   	PID: 60034\\
 4.5\,$\mu$m          	&  IRAC/{\it Spitzer}  		&   	PID: 60034\\
 5.8\,$\mu$m          	&  IRAC/{\it Spitzer}   	&   	PID: 137\\
 8.0\,$\mu$m          	&  IRAC/{\it Spitzer}   	&   	PID: 137\\
 450/850\,$\mu$m    &  SCUBA-2/JCMT    		& 	This work  \\
 20~cm                	&  VLA   				& 	Wold et al. 2012            \\
\end{tabular}
}
\label{mw}
\end{center}
\end{table}

\section{Source Identification and Catalog}

\subsection{Source Extraction}

On both maps, we first extracted sources that have a peak S/N greater than 3.0 within the effective regions. The way we extracted the source is that we found the maximum pixel within the desired region, took the position and the value information of the peak, and subtracted a PSF centered at the source and scaled accordingly. We iterated this process until the peak S/N value was less than 3.0. We treated these peaks as the preliminary catalogs at 450\,$\mu$m and 850\,$\mu$m. We then cross-correlated the two catalogs to find the counterparts in the other bandpass. We considered a counterpart recovered if the position of the 450\,$\mu$m source lay within the 850\,$\mu$m beam. The only exception was A370-450.9, which lies slightly outside the beam area of A370-850.5 but likely contributes a certain amount of the 850\,$\mu$m flux and makes A370-850.5 appear elongated. We considered both A370-450.9 and A370-450.11 to be counterparts to A370-850.5.
  
To form the final catalogs, we kept every $>4~\sigma$ source in the preliminary catalogs, as well as every $>3~\sigma$ source that had a $>3~\sigma$ counterpart in the other bandpass. In Table~\ref{450} (\ref{850}), we give the $>4~\sigma$ 450 (850)\,$\mu$m sources, as well as the $>3~\sigma$ 450 (850)\,$\mu$m sources with $>3~\sigma$ 850 (450)\,$\mu$m counterparts. In Column~1, we give the name of the source; in Column~2, our identification for the source; in Columns~3 and 4, the right ascension and declination in J2000 coordinates; in Column~5, the S/N in the discovery bandpass; in Column~6, the flux in the discovery bandpass;  in Column~7, our identification for the counterpart in the other bandpass, where available; in Column~8, the S/N for that counterpart or the S/N measured at the peak position in the other bandpass; in Column~9, the flux for that counterpart or the flux measured at the peak position in the other bandpass.  

\begin{figure*}[h]
 \begin{center}
    \leavevmode
      \includegraphics[width=0.77 \textwidth]{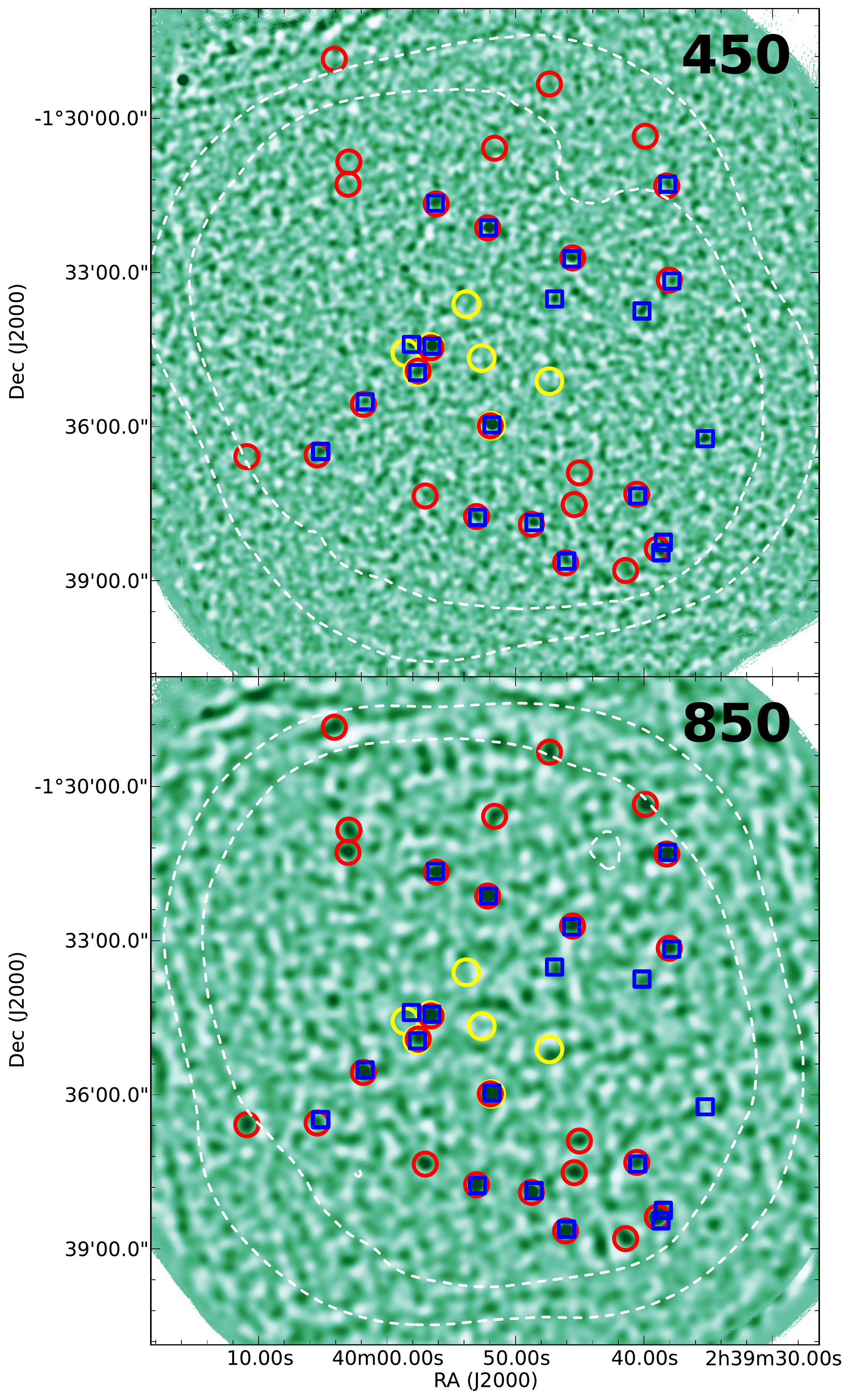}
       \caption{A370 SCUBA-2 S/N maps at 450\,$\mu$m (top panel) and 850\,$\mu$m (bottom panel) with a linear green scale from -4 to 4. The size of both maps is 13$' \times$ 13$'$. Blue squares are the 20 450\,$\mu$m sources given in Table~\ref{450}, and red circles are the 26 850\,$\mu$m sources given in Table~\ref{850}. Yellow circles are SCUBA 850\,$\mu$m sources from \citet{Cowie:2002p2075}. White dashed curves are noise contours with levels 0.82 $\times$ (2, 3) mJy/beam at 850\,$\mu$m and 3.92 $\times$ (2, 3) mJy/beam at 450\,$\mu$m. The sizes of the circles and squares correspond to 2 times the beam FWHM of their respective wavelength (Table \ref{cal}).}
     \label{a370850450}
  \end{center}
\end{figure*}

We give our assessment of the reliability of our source extraction in Section~3.2. We show our final S/N images in Figure~\ref{a370850450} overplotted with all the sources in Tables~\ref{450} and \ref{850} (red circles and blue squares for 850\,$\mu$m and 450\,$\mu$m, respectively) and with the noise contours. In total, we detected 26 sources at 850\,$\mu$m and 20 sources at 450\,$\mu$m. The number density is 3 (10) sources per 1000 beam areas at 450 (850)\,$\mu$m, which indicates that our maps are not yet reaching the confusion limit (roughly 1 source per 30 beam areas). Thus, the flux boosting and the positional errors caused by source confusion should be negligible (\citealt{Hogg:2001uq}).  Note that the maps are all gridded into 1$''$ by 1$''$ pixels. We experimented with different pixel sizes and found that the results are not affected by the different choices. This is because the matched-filter technique is not affected by the pixel size (Equations~1 and 2), as long as the PSF is well sampled.

\begin{table*}
\begin{center}
\caption{SCUBA-2 450 micron detected sources}
\scalebox{1}{
\begin{tabular}{llllrclcc}
\hline
 Name                  &  ID           &  R.A.        &  DEC       &    S/N  &  $f_{450}$         &   Counterpart  &  S/N$_{c,850}$  &  $f_{c,850}$         \\
                       &               &  (J2000)     &  (J2000)     &         &  (mJy)               &      ID$_{850}$         &          (mJy)  &                    \\
                       (1)&	(2)&	(3)&	(4)&	(5) & (6)&(7)&(8)&(9) \\
\hline
 \multicolumn{9}{l}{ $>4~\sigma$ Sample} \\ 
 SMM J023951.8-013558  &  A370-450.1   &  2 39 51.83  &  -1 35 58.0  &  11.97  &  54.40 $\pm$ 4.54  &  A370-850.1   &        20.48  &  20.40 $\pm$ 1.00  \\
 SMM J023956.5-013426  &  A370-450.2   &  2 39 56.50  &  -1 34 26.0  &  10.93  &  45.81 $\pm$ 4.19  &  A370-850.3   &         9.16  &  7.89 $\pm$ 0.86   \\
 SMM J023952.1-013208  &  A370-450.3   &  2 39 52.10  &  -1 32 08.0  &   8.79  &  46.63 $\pm$ 5.31  &  A370-850.2   &        11.68  &  11.93 $\pm$ 1.02  \\
 SMM J023945.6-013244  &  A370-450.4   &  2 39 45.63  &  -1 32 44.0  &   5.18  &  31.17 $\pm$ 6.02  &  A370-850.18  &         4.18  &  4.96 $\pm$ 1.19   \\
 SMM J023948.6-013752  &  A370-450.5   &  2 39 48.56  &  -1 37 52.0  &   4.76  &  26.09 $\pm$ 5.48  &  A370-850.6   &         7.35  &  8.45 $\pm$ 1.15   \\
 SMM J023935.2-013614  &  A370-450.6   &  2 39 35.23  &  -1 36 14.0  &   4.69  &  33.48 $\pm$ 7.13  &  \ldots       &        -1.04  &  -1.48 $\pm$ 1.43  \\
 SMM J023953.0-013746  &  A370-450.7   &  2 39 52.97  &  -1 37 46.0  &   4.59  &  26.29 $\pm$ 5.73  &  A370-850.8   &         6.06  &  6.98 $\pm$ 1.15   \\
 SMM J023947.0-013331  &  A370-450.8   &  2 39 46.96  &  -1 33 31.0  &   4.50  &  21.21 $\pm$ 4.72  &  \ldots       &         2.43  &  2.38 $\pm$ 0.98   \\
 SMM J023938.5-013815  &  A370-450.9   &  2 39 38.49  &  -1 38 15.0  &   4.46  &  29.30 $\pm$ 6.57  &  A370-850.5   &         7.43  &  11.13 $\pm$ 1.50  \\
 SMM J023958.1-013424  &  A370-450.10  &  2 39 58.10  &  -1 34 24.0  &   4.38  &  19.45 $\pm$ 4.44  &  \ldots       &         1.62  &  1.48 $\pm$ 0.91   \\
 SMM J023938.7-013827  &  A370-450.11  &  2 39 38.69  &  -1 38 27.0  &   4.05  &  26.73 $\pm$ 6.60  &  A370-850.5   &         7.43  &  11.13 $\pm$ 1.50  \\
 SMM J023940.2-013345  &  A370-450.12  &  2 39 40.16  &  -1 33 45.0  &   4.03  &  21.98 $\pm$ 5.46  &  \ldots       &         1.90  &  2.17 $\pm$ 1.14   \\
 \multicolumn{9}{l}{ $>3~\sigma$ Sample with $>3~\sigma$ Counterparts at 850\,$\mu$m} \\ 
SMM J024005.2-013629  &  A370-450.13  &  2 40 05.17  &  -1 36 29.0  &   3.79  &  21.82 $\pm$ 5.75  &  A370-850.26  &         3.07  &  4.27 $\pm$ 1.39   \\
 SMM J023957.6-013457  &  A370-450.14  &  2 39 57.64  &  -1 34 57.0  &   3.77  &  16.80 $\pm$ 4.46  &  A370-850.17  &         4.21  &  3.77 $\pm$ 0.90   \\
 SMM J023946.0-013837  &  A370-450.15  &  2 39 46.03  &  -1 38 37.0  &   3.75  &  22.88 $\pm$ 6.10  &  A370-850.9   &         5.74  &  7.42 $\pm$ 1.29   \\
 SMM J023956.2-013139  &  A370-450.16  &  2 39 56.23  &  -1 31 39.0  &   3.67  &  19.25 $\pm$ 5.24  &  A370-850.10  &         5.58  &  6.05 $\pm$ 1.08   \\
 SMM J023938.2-013117  &  A370-450.17  &  2 39 38.16  &  -1 31 17.0  &   3.41  &  28.24 $\pm$ 8.27  &  A370-850.4   &         7.48  &  11.79 $\pm$ 1.58  \\
 SMM J024001.7-013531  &  A370-450.18  &  2 40 01.70  &  -1 35 31.0  &   3.19  &  16.24 $\pm$ 5.09  &  A370-850.7   &         6.96  &  7.61 $\pm$ 1.09   \\
 SMM J023940.5-013721  &  A370-450.19  &  2 39 40.50  &  -1 37 21.0  &   3.15  &  20.74 $\pm$ 6.59  &  A370-850.25  &         3.88  &  5.28 $\pm$ 1.36   \\
 SMM J023937.8-013310  &  A370-450.20  &  2 39 37.83  &  -1 33 10.0  &   3.12  &  18.94 $\pm$ 6.08  &  A370-850.15  &         4.08  &  5.14 $\pm$ 1.26   \\
\hline
\end{tabular}
}
\label{450}
\end{center}
\end{table*}

\begin{table*}
\begin{center}
\caption{SCUBA-2 850 micron detected sources}
\scalebox{1}{
\begin{tabular}{clllrclcc}
\hline
 Name                  &  ID           &  R.A.        &  DEC       &    S/N  &  $f_{850}$         &  Counterpart    &  S/N$_{\mathrm{c,450}}$  &  $f_{c,450}$       \\
                       &               &  (J2000)     &  (J2000)     &         &  (mJy)               &       ID$_{450}$        &                     (mJy)  &                    \\
                       (1)&	(2)&	(3)&	(4)&	(5) & (6)&(7)&(8)&(9) \\
\hline
 \multicolumn{9}{l}{ $>4~\sigma$ Sample} \\ 
 SMM J023952.0-013559  &  A370-850.1   &  2 39 51.97  &  -1 35 59.0  &  20.48  &  20.40 $\pm$ 1.00  &  A370-450.1   &        11.97  &  54.40 $\pm$ 4.54  \\
 SMM J023952.2-013208  &  A370-850.2   &  2 39 52.17  &  -1 32 08.0  &  11.68  &  11.93 $\pm$ 1.02  &  A370-450.3   &         8.79  &  46.63 $\pm$ 5.31  \\
 SMM J023956.6-013428  &  A370-850.3   &  2 39 56.57  &  -1 34 28.0  &   9.16  &  7.89 $\pm$ 0.86   &  A370-450.2   &        10.93  &  45.81 $\pm$ 4.19  \\
 SMM J023938.2-013119  &  A370-850.4   &  2 39 38.23  &  -1 31 19.0  &   7.48  &  11.79 $\pm$ 1.58  &  A370-450.17  &         3.41  &  28.24 $\pm$ 8.27  \\
 SMM J023939.0-013823  &  A370-850.5   &  2 39 38.96  &  -1 38 23.0  &   7.43  &  11.13 $\pm$ 1.50  &  A370-450.11  &         4.05  &  26.73 $\pm$ 6.60  \\
                       &               &              &              &         &                    &  A370-450.9   &         4.46  &  29.30 $\pm$ 6.57  \\
 SMM J023948.8-013754  &  A370-850.6   &  2 39 48.76  &  -1 37 54.0  &   7.35  &  8.45 $\pm$ 1.15   &  A370-450.5   &         4.76  &  26.09 $\pm$ 5.48  \\
 SMM J024001.8-013534  &  A370-850.7   &  2 40 01.84  &  -1 35 34.0  &   6.96  &  7.61 $\pm$ 1.09   &  A370-450.18  &         3.19  &  16.24 $\pm$ 5.09  \\
 SMM J023953.0-013745  &  A370-850.8   &  2 39 53.03  &  -1 37 45.0  &   6.06  &  6.98 $\pm$ 1.15   &  A370-450.7   &         4.59  &  26.29 $\pm$ 5.73  \\
 SMM J023946.1-013839  &  A370-850.9   &  2 39 46.10  &  -1 38 39.0  &   5.74  &  7.42 $\pm$ 1.29   &  A370-450.15  &         3.75  &  22.88 $\pm$ 6.10  \\
 SMM J023956.2-013140  &  A370-850.10  &  2 39 56.17  &  -1 31 40.0  &   5.58  &  6.05 $\pm$ 1.08   &  A370-450.16  &         3.67  &  19.25 $\pm$ 5.24  \\
 SMM J024003.0-013117  &  A370-850.11  &  2 40 03.04  &  -1 31 17.0  &   5.27  &  6.98 $\pm$ 1.33   &  \ldots       &         2.14  &  12.41 $\pm$ 5.79  \\
 SMM J023957.0-013721  &  A370-850.12  &  2 39 57.03  &  -1 37 21.0  &   5.26  &  6.79 $\pm$ 1.29   &  \ldots       &         2.03  &  14.03 $\pm$ 6.91  \\
 SMM J023939.9-013021  &  A370-850.13  &  2 39 39.90  &  -1 30 21.0  &   5.18  &  8.43 $\pm$ 1.63   &  \ldots       &         0.26  &  2.31 $\pm$ 8.83   \\
 SMM J023945.4-013731  &  A370-850.14  &  2 39 45.43  &  -1 37 31.0  &   4.72  &  5.72 $\pm$ 1.21   &  \ldots       &         1.05  &  6.08 $\pm$ 5.80   \\
 SMM J023938.0-013309  &  A370-850.15  &  2 39 38.03  &  -1 33 09.0  &   4.60  &  5.77 $\pm$ 1.25   &  A370-450.20  &         3.12  &  18.94 $\pm$ 6.08  \\
 SMM J023941.4-013848  &  A370-850.16  &  2 39 41.43  &  -1 38 48.0  &   4.33  &  6.40 $\pm$ 1.48   &  \ldots       &         1.95  &  12.81 $\pm$ 6.58  \\
 SMM J023957.6-013455  &  A370-850.17  &  2 39 57.57  &  -1 34 55.0  &   4.21  &  3.77 $\pm$ 0.90   &  A370-450.14  &         3.77  &  16.80 $\pm$ 4.46  \\
 SMM J023945.6-013243  &  A370-850.18  &  2 39 45.56  &  -1 32 43.0  &   4.18  &  4.96 $\pm$ 1.19   &  A370-450.4   &         5.18  &  31.17 $\pm$ 6.02  \\
 SMM J023947.4-012920  &  A370-850.19  &  2 39 47.36  &  -1 29 20.0  &   4.13  &  6.89 $\pm$ 1.67   &  \ldots       &         1.77  &  15.52 $\pm$ 8.78  \\
 SMM J024004.1-012851  &  A370-850.20  &  2 40 04.10  &  -1 28 51.0  &   4.10  &  8.32 $\pm$ 2.03   &  \ldots       &         0.22  &  3.39 $\pm$ 15.37  \\
 SMM J024010.9-013635  &  A370-850.21  &  2 40 10.91  &  -1 36 35.0  &   4.04  &  6.93 $\pm$ 1.72   &  \ldots       &        -0.10  &  -0.76 $\pm$ 7.64  \\
 SMM J024003.0-013051  &  A370-850.22  &  2 40 02.97  &  -1 30 51.0  &   4.03  &  5.22 $\pm$ 1.30   &  \ldots       &         0.57  &  3.40 $\pm$ 5.93   \\
 SMM J023945.0-013654  &  A370-850.23  &  2 39 45.03  &  -1 36 54.0  &   4.02  &  5.21 $\pm$ 1.30   &  \ldots       &         1.59  &  9.52 $\pm$ 6.00   \\
 SMM J023951.6-013035  &  A370-850.24  &  2 39 51.63  &  -1 30 35.0  &   4.00  &  4.64 $\pm$ 1.16   &  \ldots       &         1.33  &  8.38 $\pm$ 6.28   \\
 \multicolumn{9}{l}{ $>3~\sigma$ Sample with $>3~\sigma$ Counterparts at 450\,$\mu$m} \\ 
SMM J023940.6-013719  &  A370-850.25  &  2 39 40.56  &  -1 37 19.0  &   3.88  &  5.28 $\pm$ 1.36   &  A370-450.19  &         3.15  &  20.74 $\pm$ 6.59  \\
 SMM J024005.4-013633  &  A370-850.26  &  2 40 05.44  &  -1 36 33.0  &   3.07  &  4.27 $\pm$ 1.39   &  A370-450.13  &         3.79  &  21.82 $\pm$ 5.75  \\
\hline
\end{tabular}
}
\label{850}
\end{center}
\end{table*}

\subsection{Reliability of Source Extraction}

We used two methods to quantify the number of spurious sources above 4\,$\sigma$ that might be included in our catalogs. First, we created a map with the same spatial dimensions as the real maps, populated the pixels with random Gaussian signal convolved with the PSF at the given wavelength, and then ran the peak identification. We iterated this process 100 times at each wavelength and found that we should expect 1.8$\pm$1.3 sources at 450\,$\mu$m and 0.6$\pm$0.8 sources at 850\,$\mu$m to be spurious. 

Second, we performed our same source extraction method on the inverted maps, where the negative signal becomes positive and vice versa, to see how many negative sources are present at $>4~\sigma$. We found 4 and 1 at 450 and 850\,$\mu$m, respectively, which agrees reasonably well with our first approach and means that we could have that  many spurious sources in our $>4~\sigma$ catalogs. 

For the 3\,$\sigma$ sources identified as being present in both bandpasses, we again created random noise maps as described above, extracted sources between 3 and 4\,$\sigma$, and then cross-correlated them with the $>3~\sigma$ sources extracted from the real maps at the other bandpass. We iterated this process 1000 times and found 0.2 and 0.3 spurious sources at 450 and 850\,$\mu$m. In other words, sources identified at the 3\,$\sigma$ level at both wavelengths are even more reliable than $>4~\sigma$ sources only detected in one bandpass.

\subsection{Comparison with Previous Catalogs}

A370 was surveyed by SCUBA at both 450 and 850\,$\mu$m within the circular region centered at the cluster centroid with a radius less than 1$\farcm$5 (\citealt{Cowie:2002p2075, Smail:2002p6793}). A370-850.3 was also observed (\citealt{Chen:2011p11605}) with the high-resolution Submillimeter Array (SMA; \citealt{Ho:2004p8376}). We list our detected sources within the SCUBA observed region along with previous results in Table \ref{compsrc850}. Our maps do not recover all the sources detected in \citet{Cowie:2002p2075}, because their maps have better sensitivity ($<0.8$~mJy/beam) at 850\,$\mu$m. However, most of the recovered sources agree with the literature in terms of their fluxes and positions, except for the flux on A370-450.1 given by \citet{Smail:2002p6793}. \citet{Knudsen:2008p3824} found that it is difficult to reliably calibrate the SCUBA 450\,$\mu$m sources, because the PSF is not well described by a 2D Gaussian and is very sensitive to the deformation of the JCMT dish. Thus, the 450\,$\mu$m fluxes obtained from SCUBA could be problematic. This problem is significantly less in the SCUBA-2 maps, since, as we showed in Table \ref{cal}, the PSFs at both 450\,$\mu$m and 850\,$\mu$m are stable and well modeled.

\begin{table*}
\begin{center}
\caption{SCUBA-2 850 micron sources compared with the literature}
\scalebox{0.8}{
\begin{tabular}{ccccccccccc}
\hline
 \hline
 Source ID    &  S$_{850}$         &  Chen et al.\ S$_{850}$  &  R.A. offset  &  DEC offset  &  Cowie et al.\ S$_{850}$  &  R.A. offset  &  DEC offset  &  Smail et al.\ S$_{850}$  &  R.A. offset  &  DEC offset  \\
&  (mJy)             &  (mJy)                  &  ($''$)    &  ($''$)    &  (mJy)                   &    ($''$)  &    ($''$)  &  (mJy)                   &  ($''$)    &  ($''$)    \\
\hline
 A370-850.1   &  20.40 $\pm$ 1.00  &  \ldots                 &  \ldots       &  \ldots       &  21.06 $\pm$ 1.34        &         -1.0  &          0.0  &  23.0 $\pm$ 1.9          &  -1.0         &  0.0          \\
 A370-850.3   &  7.89 $\pm$ 0.86   &  7.95 $\pm$ 0.60        &  -0.3         &  -1.5         &  6.68 $\pm$ 0.58         &          0.9  &         -1.0  &  11.0 $\pm$ 1.9          &  -2.6         &  -1.0         \\
 A370-850.17  &  3.77 $\pm$ 0.90   &  \ldots                 &  \ldots       &  \ldots       &  3.49 $\pm$ 0.66         &          1.0  &          1.0  &  \ldots                  &  \ldots       &  \ldots       \\
\hline
\end{tabular}
}
\label{compsrc850}
\end{center}
\end{table*}

\begin{table}
\begin{center}
\caption{SCUBA-2 450 micron sources compared with the literature}
\scalebox{0.85}{
\begin{tabular}{ccccc}
\hline
\hline
 Source ID   &  S$_{450}$         &  Smail et al. S$_{450}$  &  R.A. offset  &  DEC. offset  \\
             &  (mJy)             &  (mJy)                   &    ($''$)   &    ($''$)   \\
\hline
 A370-450.1   &  54.40 $\pm$ 4.54  &  85 $\pm$ 10             &  1.0          &  1.0          \\
 A370-450.2   &  45.81 $\pm$ 4.19  &  42 $\pm$ 10             &  -1.5         &  1.0          \\
\hline
\end{tabular}
}
\label{compsrc450}
\end{center}
\end{table}

\subsection{Completeness Tests}

\begin{figure}
 \begin{center}
    \leavevmode
      \includegraphics[scale=0.32]{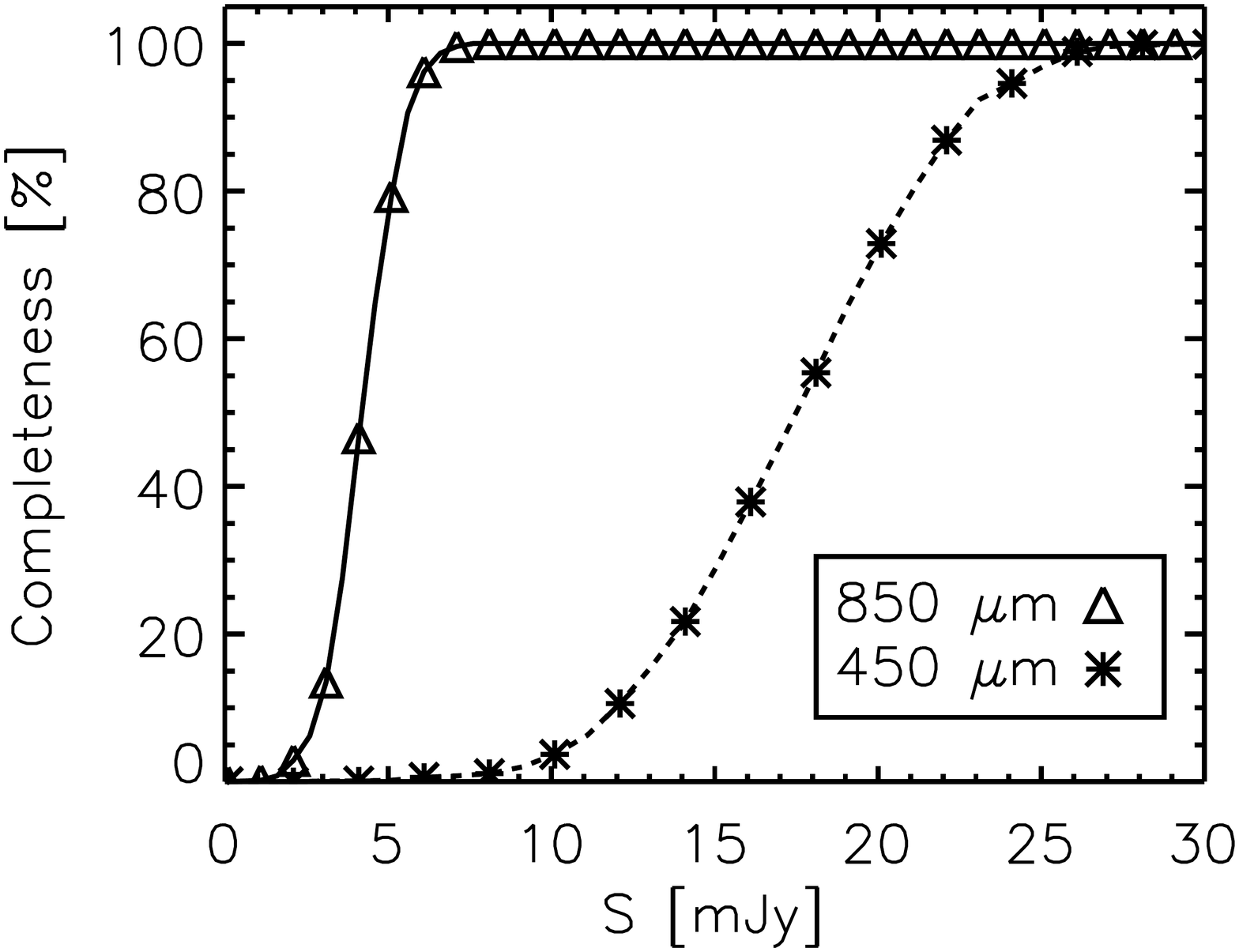}
       \caption{The result of the completeness tests on A370 at 450\,$\mu$m (asterisks, dashed curve) and 850\,$\mu$m (triangles, solid curve) vs. flux. In the simulation, the flux step at 450 (850)\,$\mu$m is 1~mJy (0.5~mJy). For a clear visual demonstration, the spacing between the asterisks is 2~mJy, and the spacing between the triangles is 1~mJy.}
     \label{completeness}
  \end{center}
\end{figure}

We took the noise maps created in the previous sections and ran a completeness test. We randomly populated the image with sources of a given flux and then ran the source extraction on the maps down to 4\,$\sigma$ to see if we could recover the added sources. A source is considered recovered if it is detected above 4\,$\sigma$ and if its position is within the beam area. We iterated this process 5000 times on each map with a flux step of 1 mJy (0.5 mJy) from 0.1 to 30.1 mJy for 450 (850)\,$\mu$m to estimate the completeness. We plot the results in Figure \ref{completeness}. The 50\% completeness is around 4 and 17~mJy at 850 and 450\,$\mu$m, respectively, and the 80\% completeness is around 5 and 21~mJy, respectively.

\section{Number Counts}

A370 is one of the most extensively studied gravitational lensing cluster fields with a well-constrained lensing model. We used LENSTOOL, a software package that models the effects of gravitational lensing by taking three-dimensional mass distributions within the cluster into account (\citealt{Kneib:1996p3751}), and adopted the latest lensing model from \citet{Richard:2010fk} to compute the lensing magnification. We first derived raw number counts directly from the detected sources in our catalogs. We then estimated the true number counts and the errors on the raw number counts using Monte Carlo simulations. Finally, we calculated the amount of EBL resolved based on the final adopted true number counts curves. We discuss our procedure in more detail in the following subsections. Note that in order to undertake a simple and clear analysis, in this section we only used sources that were independently and uniformly selected at $>4~\sigma$ in each bandpass without regard to whether there was a submillimeter counterpart in the other bandpass.

\subsection{Delensed Raw Number Counts}

The SCUBA-2 850\,$\mu$m sources have a coarse resolution with a beam size of 14$''$, which can cause large uncertainties in estimating the magnifications where the sources are strongly amplified (\citealt{Chen:2011p11605}). Thus, for 850\,$\mu$m sources with a clear counterpart identified in either deep radio interferometric maps, 450\,$\mu$m data, or submillimeter interferometric maps from the SMA (see Section~5), we adopted the redshifts and the positions of the counterparts for estimating the magnifications. The A370 cluster itself is at $z = 0.37$. For any sources having moderate amplifications with redshifts beyond $z = 1$, the amplification is only weakly dependent on the redshifts (\citealt{Blain:1999p7279}). Thus, for sources without redshift measurements, we adopt $z = 2.5$ based on the current finding that the median redshift of the SMG population is around 2.5 (\citealt{Chapman:2005p5778, Wardlow:2011qy}).

Number counts are usually described in two ways: cumulative number counts or differential number counts. When the sample size is small, cumulative number counts give a cleaner visual picture of the underlying shape of the counts; however, errors on the cumulative counts are correlated and need to be carefully taken into account when one tries to fit the counts. On the other hand, differential counts have independent errors on each count measurement, making it straightforward to do $\chi^2$ fitting on the counts. Our sample size is big enough with at least 10 sources in each bandpass that we can focus on analyzing the differential counts. In some cases we will also present the cumulative counts, but those are based on the differential counts. 

In a blank-field survey, the differential number counts can simply be calculated by dividing each source in a given flux bin by the area surveyed, summing these measurements, and then dividing by the flux bin width. However, in massive cluster fields, because of the gravitational lensing, the area used to divide each source is not simply the image plane area. It should be the total source plane area in which each source is located and could be significantly detected with its flux. 

\begin{figure}
 \begin{center}
    \leavevmode
      \includegraphics[scale=0.33]{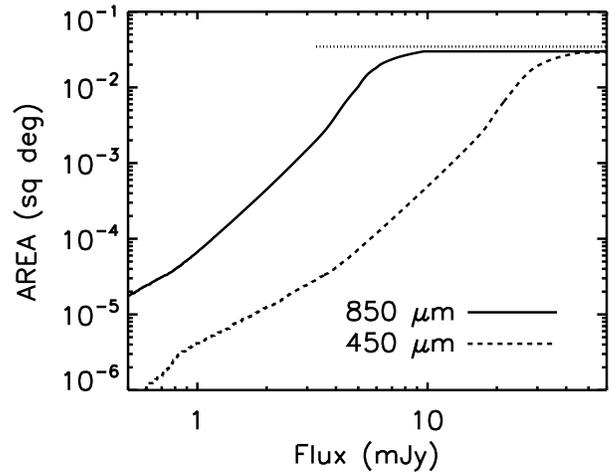}
       \caption{At $z=2.5$, the source plane areas over which a source with a given flux in the source plane would be detected at $>4~\sigma$ in the image plane. Dashed (solid) curves are for 450 (850)\,$\mu$m. For comparison, we also show the 850\,$\mu$m image plane area as the horizontal dotted segment with the lower flux limit representing four times the 850\,$\mu$m r.m.s.\ at the center of the image.}
     \label{spf2p5}
  \end{center}
\end{figure}

To find the detectable source plane area for each source, we gridded the image plane into 1$''$ pixels and projected it back to the source plane at the source redshift. The total image plane area is the sum of any part of the map that has more than 1/9 of the central exposure time (Table \ref{obs}). We found the spatial boundary of the source plane and then calculated the total source plane area. The average magnification of our observations at a given redshift is the ratio between the image plane area and the source plane area, which is 1.16 at both wavelengths at $z = 2.5$. We then regridded the source plane back into 1$''$ pixels and projected them to the image plane. We counted the number of source grid points that were detectable at $>4~\sigma$ on the image plane, assuming each source grid emits the specified flux, and then multiplied the total number of points by 1~arcsec$^2$ to obtain the total detectable source plane area. In Figure~\ref{spf2p5}, we demonstrate the source plane areas over which a source with a given flux in the source plane could be detected at $>4.0~\sigma$ in the image plane, assuming $z = 2.5$.

\begin{figure}[h]
 \begin{center}
    \leavevmode
      \includegraphics[scale=0.7]{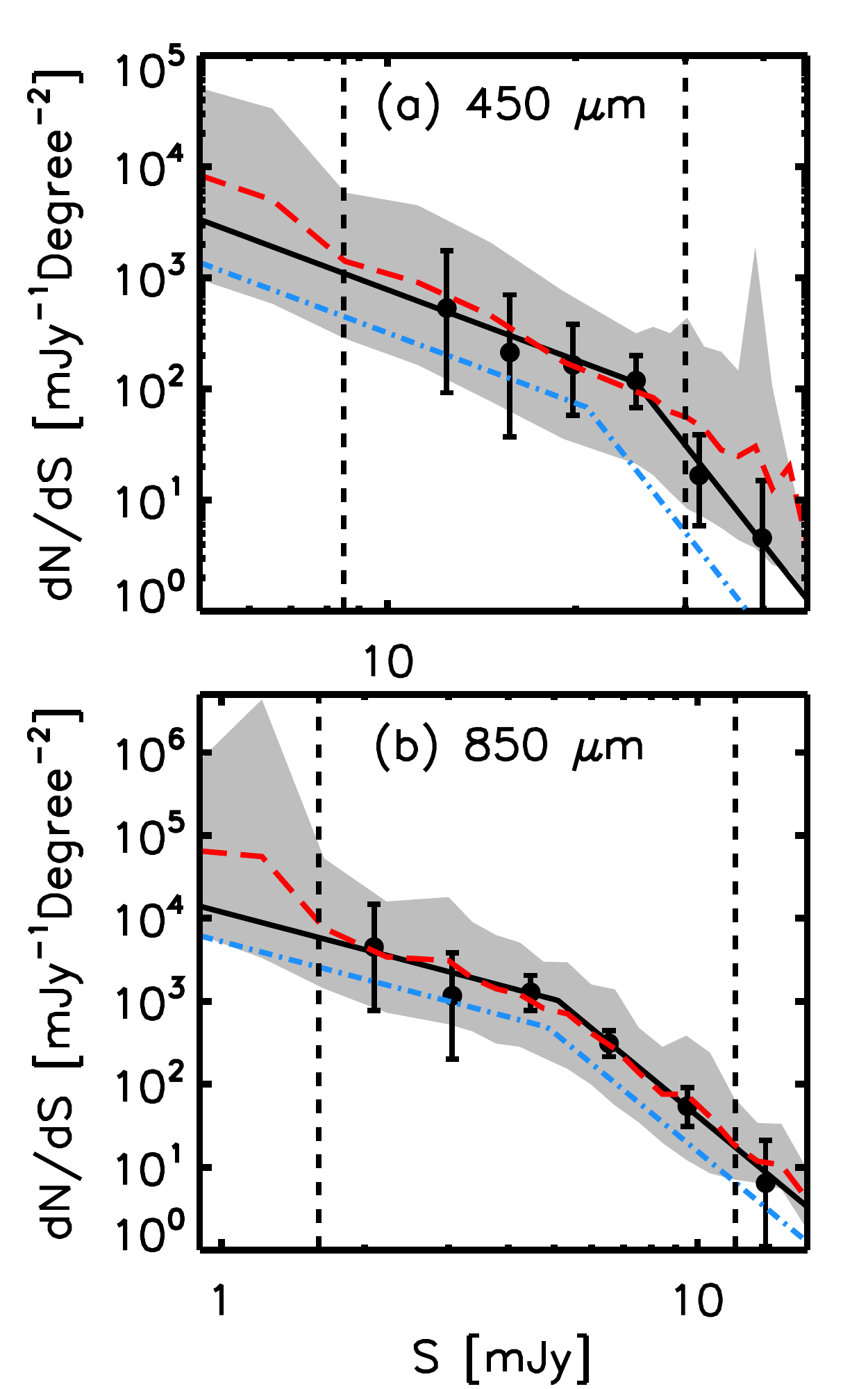}
       \caption{The raw and simulated differential number counts at (a) 450\,$\mu$m and (b) 850\,$\mu$m. In each panel, the black circles are the raw number counts, and the black curves are the $\chi^2$ broken power law fits to the raw number counts. The blue dot-dashed curves, representing the true number counts, are the final adopted model curves for our Monte Carlo simulations (see Section~4.2). The red dashed curves and the shaded regions are, respectively, the recovered mean number counts and the 90\% confidence range obtained from the Monte Carlo simulations. The flux ranges enclosed by the black dashed lines are where the recovered curves are well constrained.}
     \label{errcouts}
  \end{center}
\end{figure}

We inverted the source plane area of each individual source to obtain the source surface density. Then, we added up the surface density of sources lying within each flux bin and divided that sum by the flux bin width to obtain the differential counts. 
The differential counts are best described by a broken power law, because once the map is sensitive enough to resolve the bulk of the EBL, the faint-end slope of the counts needs to be shallower than the bright-end slope so that the total EBL is finite. We performed $\chi^2$ fits using the broken power law with basically two different slopes on each side of the characteristic flux. The sources with the characteristic flux dominate the contribution to the EBL. The broken power law is of the form
\begin{equation}
  \frac{dN}{dS} = \left\{
  \begin{array}{l l}
    {N_0}\left(\frac{S}{S_0}\right)^{-\alpha}  & \quad \text{if $S \leq S_0$}\\
    {N_0}\left(\frac{S}{S_0}\right)^{-\beta}  & \quad \text{if $S > S_0$}\\
  \end{array} \right..
\end{equation} 
We show the results of our fits in Table \ref{fr}. We also plot them in Figure \ref{errcouts} (black lines), along with the raw number counts (black circles). Given the complicated effects of gravitational lensing, uneven sensitivity, and confusion, the true number counts and the errors on the raw number counts are best estimated with Monte Carlo simulations.

\begin{figure}[h]
 \begin{center}
    \leavevmode
      \includegraphics[scale=0.7]{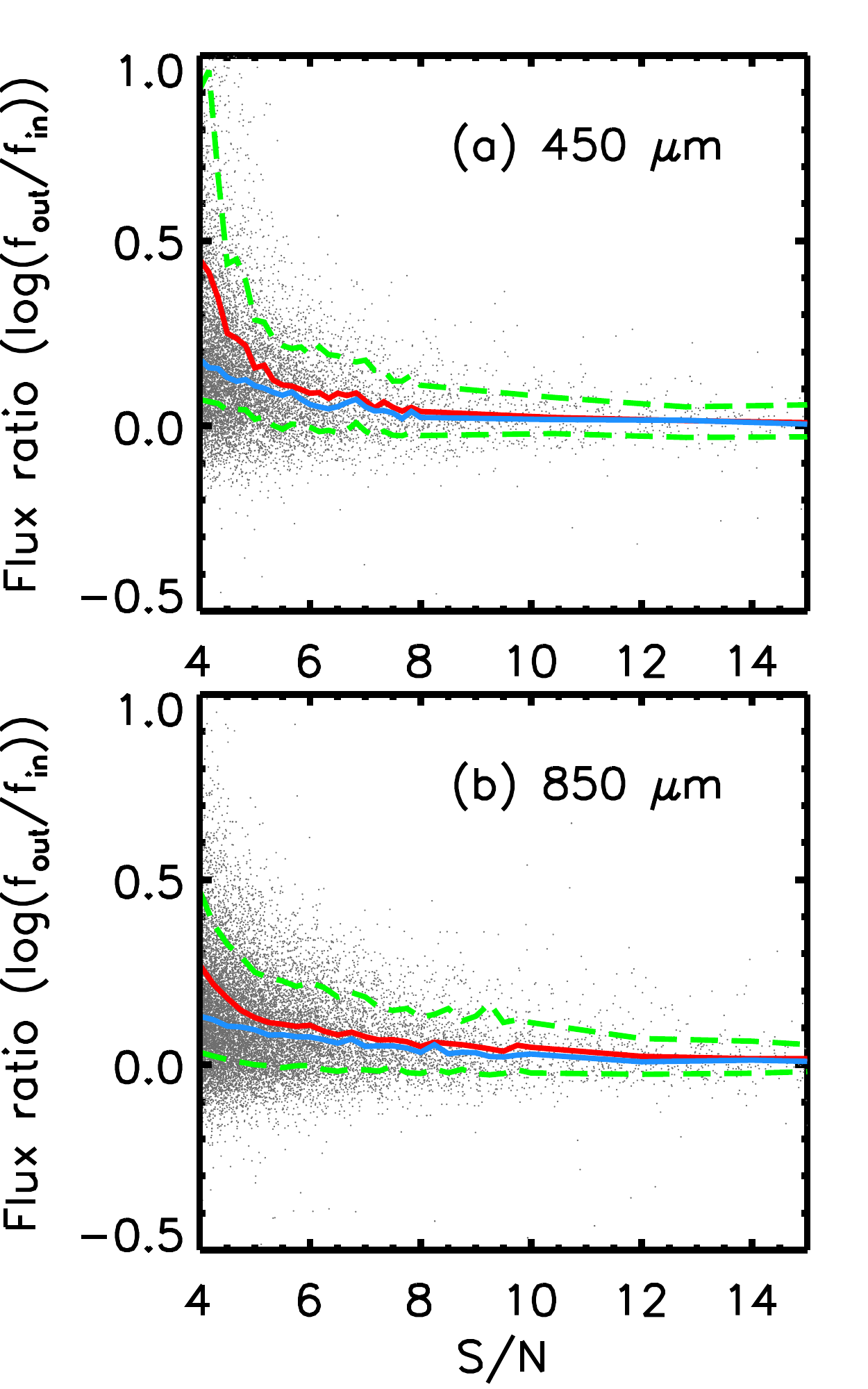}
       \caption{The ratio between the delensed fluxes of the detected sources and the input fluxes from our Monte Carlo simulations. Black dots are $\sim$10000 simulated data points. Red (blue) curves are the mean (median) value of the flux ratio in the different S/N bins. The area enclosed by the green dashed curves represents the 1\,$\sigma$ range relative to the mean values.}
     \label{flxbst}
  \end{center}
\end{figure}

\begin{figure*}[t]
 \begin{center}
      \includegraphics[scale=1.0]{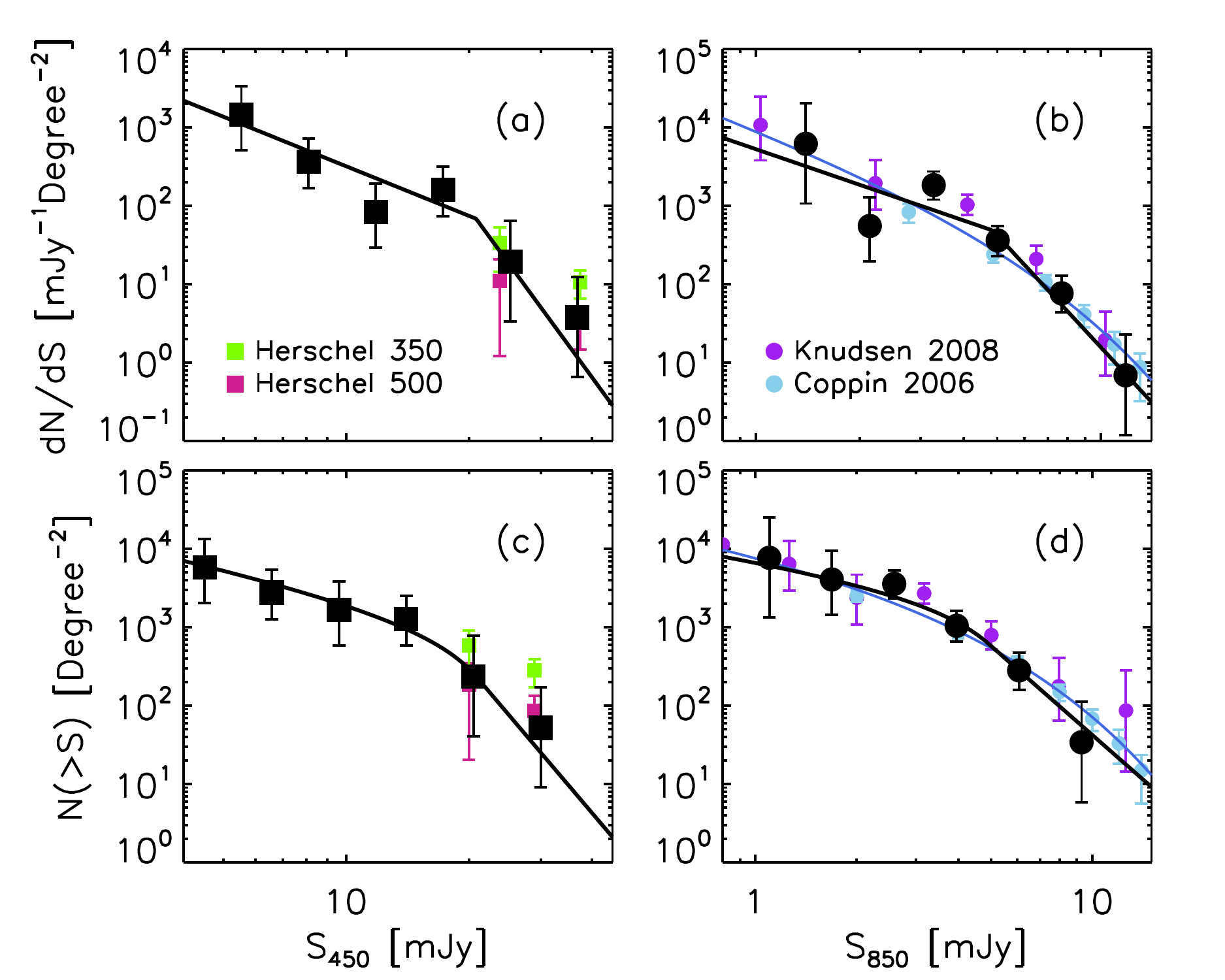}
       \caption{SCUBA-2 A370 differential number counts at (a) 450\,$\mu$m and (b) 850\,$\mu$m and cumulative number counts at (c) 450\,$\mu$m and (d) 850\,$\mu$m. Black filled symbols (squares for 450\,$\mu$m and circles for 850\,$\mu$m) are the delensed and deboosted number counts given in Tables~\ref{cn4} and \ref{cn8}. In all panels, the black curves are our true number counts curves (i.e., the blue dot-dashed curves from Figure~\ref{errcouts}). Number counts obtained from {\it Herschel} at 350\,$\mu$m and 500\,$\mu$m  (\citealt{Oliver:2010p11204}) are denoted by olive and pink squares, respectively, in panels (a) and (c). Blue curves in (b) and (d) show the 850\,$\mu$m fit from \citet{Knudsen:2008p3824}. Their number counts are denoted by purple filled circles. The blank-field counts from SHADES (\citealt{Coppin:2006p9123}) are denoted by blue filled circles. }
     \label{counts}
  \end{center}
\end{figure*}

\begin{table}
\begin{center}
\caption{$\chi^2$ fits on the 450 and 850 micron raw differential number counts}
\begin{tabular}{ccrrrr}
\hline
\hline
 Wavelengths  &  Equation  &  N$_0$  &  S$_0$  &  $\alpha$  &  $\beta$  \\
\hline
         450  &  (3)       &    114  &  25.0  &      2.10  &     7.12  \\
         850  &  (3)       &    1024  &   5.09  &      1.51  &     4.75  \\
\hline
\end{tabular}
\label{fr}
\end{center}
\end{table}

\subsection{Simulations}

We first estimated the true noise using the Jackknife resampling technique by subtracting two maps that were each created from coadding roughly half of the data. Since the real sources, regardless of the significance of the detection, are subtracted off, the residual maps are source-free real noise maps. We then estimated the true noise from the noise maps scaled by a factor of $\sqrt{t1\times t2}$/(t1+t2), with t1 and t2 representing the exposure time of each pixel from the two maps. We found central noise values of 3.82 and 0.72~mJy/beam at 450\,$\mu$m and 850\,$\mu$m, respectively, about 0.1~mJy/beam lower than the original maps containing faint sources. 

We then took the broken power law fits from Section~4.1 as the underlying models of the source populations, drew sources from these models, and populated them onto the source plane at $z=2.5$ at random positions with no clustering. For bright sources, where the average number of input sources is less than one, we randomized their input as either one or zero, with the number of each being determined by the probability of seeing a source at that flux.  We imaged the sources back onto the image plane using LENSTOOL, added them into the true noise maps, and then ran through the same procedure of source extraction and number counts calculation that we performed on the data. Note that because DIMM performs a high-pass filtering in Fourier space to filter out the low spatial frequency structures, our data maps are not sensitive to the zeroth order scaling factor seen in our simulated maps due to many evenly distributed faint sources. We therefore applied the filter on the simulated maps before we proceeded to extract sources.  

We iterated this procedure 500 times using at least 10000 simulated data points each time and compared the recovered curves from the simulations that connected the mean values of each flux bin with the models. We found a systematic flux/count boost, which we investigated by tracing the input source properties of each detection. For each $>4~\sigma$ source in the output maps, we identified the brightest input source located within the beam area as the counterpart and plotted its flux ratio in Figure~\ref{flxbst}. We did not remove potential false detections if the ratio were unreasonably high (i.e., a flux ratio $>2$), since we kept every detected source for the real counts, and it is critical to follow the same procedure for the simulated counts as for the real counts. At S/N = 4, we found median flux boosting factors of 1.61 and 1.31 at 450\,$\mu$m and 850\,$\mu$m, respectively. At 850\,$\mu$m, this is in agreement with previous SCUBA studies. \citet{Cowie:2002p2075} performed a Monte Carlo simulation with a single slope input model of the A370 field and found the boosting factor to be around 1.25 at 850\,$\mu$m. Various SCUBA blank-field studies also found similar results (\citealt{Eales:2000uq, Scott:2002p6539, Wang:2004p2270}). 

We then applied the corrections to the models by adjusting the positions of the characteristic fluxes and the scaling factors of the broken power laws. Note that we kept the measured slopes, since the boost is much more sensitive to the S/N than to the intrinsic fluxes. 

We list the parameters of our final adopted model curves (hereafter referred to as our true number counts curves) in Table~\ref{ff}, and we plot them in Figure~\ref{errcouts} (blue dot-dashed curves). We show the recovered mean number counts as the red dashed curves. We use gray shading to represent the 90\% confidence region. Between $\sim 1.6 - 12$~mJy at 850\,$\mu$m and $\sim 8.5 - 30$~mJy at 450\,$\mu$m (black dashed vertical lines), the recovered mean number counts curves agree well with the broken power law fits to the data (black solid curves) and are well constrained with at least 100 data points per bin. The fainter-end counts are overestimated by the curves at both wavelengths. Because they are not well sampled due to the small source plane area, the output counts are dominated by serendipitous detections from strongly lensed sources, and the number counts curves based on them are highly boosted. The same argument can be applied to the brighter-end counts, where the curves are more unstable and scattered. The surface density of bright sources is too low to generate an adequate number of sources given the survey area of our observations. Thus, the output counts curves are dominated by the boosted fainter sources. 

Because our maps have not yet reached the confusion limit, the cause of the boost is mostly due to the statistical fluctuations of the source flux measurements for flux-limited observations, known as the Eddington bias (\citealt{Eddington:1913fj}). Previous SCUBA studies (e.g., \citealt{Eales:2000uq}) also found that the dominant boosting factor is the noise.

\begin{table}
\begin{center}
\caption{The true number counts curves at 450 and 850 micron from the Monte Carlo simulations}
\begin{tabular}{ccrrrr}
\hline
\hline
 Wavelengths  &  Equation  &  N$_0$  &  S$_0$  &  $\alpha$  &  $\beta$  \\
\hline
         450  &  (3)       &    68.4  &  20.8  &      2.10  &     7.12  \\
         850  &  (3)       &    485  &   4.85  &      1.51  &     4.75  \\
\hline
\end{tabular}
\label{ff}
\end{center}
\end{table}

\subsection{True Number Counts and the EBL}

We took our $>4~\sigma$ sources and deboosted their fluxes based on the mean flux boosting ratio shown in Figure \ref{flxbst} (red curves). We computed the true number counts in a similar way as presented in Section~4.1. However, instead of using the delensed fluxes, we added up the surface densities of sources with delensed and deboosted fluxes lying within each flux bin. Tables \ref{cn4} and \ref{cn8} give the true differential number counts and the number of sources in each flux bin at 450\,$\mu$m and 850\,$\mu$m. The errors are based on Poisson statistics (\citealt{Gehrels:1986p1344}). The faintest fluxes we detected after delensing and deboosting are $\sim$1.1~mJy at 850\,$\mu$m and $\sim$4.5~mJy at 450\,$\mu$m.

\begin{table}
\begin{center}
\caption{450 micron True Number Counts}
\begin{tabular}{rrlrl}
\hline
 S$_{450}$  &  N  &  dN/dS$^a$                                                                       &  S$_{450}$  &  N($>$S)$^b$                                                                   \\
       (mJy)  &      &  (mJy$^{-1}$ deg$^{-2}$)                                                       &        (mJy)  &  (deg$^{-2}$)                                                                  \\
\hline
  5.54  &  2  &  $1455\begin{array}{l}\scriptstyle +1919\\\scriptstyle -939.7\end{array}$    &   4.50  &  $5786\begin{array}{l}\scriptstyle +7632\\\scriptstyle -3738\end{array}$     \\
  8.09  &  3  &  $367.5\begin{array}{l}\scriptstyle +357.5\\\scriptstyle -200.1\end{array}$  &   6.58  &  $2765\begin{array}{l}\scriptstyle +2690\\\scriptstyle -1505\end{array}$     \\
 11.83  &  2  &  $83.14\begin{array}{l}\scriptstyle +109.7\\\scriptstyle -53.71\end{array}$  &   9.61  &  $1650\begin{array}{l}\scriptstyle +2176\\\scriptstyle -1066\end{array}$     \\
 17.29  &  3  &  $161.2\begin{array}{l}\scriptstyle +156.8\\\scriptstyle -87.76\end{array}$  &  14.05  &  $1281\begin{array}{l}\scriptstyle +1246\\\scriptstyle -697.3\end{array}$    \\
 25.26  &  1  &  $19.43\begin{array}{l}\scriptstyle +44.69\\\scriptstyle -16.07\end{array}$  &  20.53  &  $236.2\begin{array}{l}\scriptstyle +543.2\\\scriptstyle -195.3\end{array}$  \\
 36.92  &  1  &  $3.76\begin{array}{l}\scriptstyle +8.65\\\scriptstyle -3.11\end{array}$     &  30.00  &  $52.09\begin{array}{l}\scriptstyle +119.8\\\scriptstyle -43.08\end{array}$  \\
\hline
\end{tabular}
\label{cn4}
\end{center}

\begin{center}
\caption{850 micron True Number Counts}
\begin{tabular}{rrlrl}
\hline
\hline
 S$_{850}$  &   N  &  dN/dS$^a$                                                                      &  S$_{850}$  &  N($>$S)$^b$                                                                   \\
       (mJy)  &      &  (mJy$^{-1}$ deg$^{-2}$)                                                       &        (mJy)  &  (deg$^{-2}$)                                                                  \\
\hline
  1.39  &  1  &  $6176\begin{array}{l}\scriptstyle +14205\\\scriptstyle -5108\end{array}$    &  1.10  &  $7683\begin{array}{l}\scriptstyle +17670\\\scriptstyle -6353\end{array}$    \\
  2.13  &  2  &  $553.0\begin{array}{l}\scriptstyle +729.4\\\scriptstyle -357.2\end{array}$  &  1.69  &  $4064\begin{array}{l}\scriptstyle +5361\\\scriptstyle -2626\end{array}$     \\
  3.27  &  8  &  $1835\begin{array}{l}\scriptstyle +905.8\\\scriptstyle -634.8\end{array}$   &  2.58  &  $3568\begin{array}{l}\scriptstyle +1762\\\scriptstyle -1235\end{array}$     \\
  5.01  &  7  &  $361.7\begin{array}{l}\scriptstyle +194.8\\\scriptstyle -133.4\end{array}$  &  3.96  &  $1044\begin{array}{l}\scriptstyle +562.1\\\scriptstyle -384.8\end{array}$   \\
  7.68  &  5  &  $76.44\begin{array}{l}\scriptstyle +51.71\\\scriptstyle -33.02\end{array}$  &  6.07  &  $281.1\begin{array}{l}\scriptstyle +190.1\\\scriptstyle -121.4\end{array}$  \\
 11.78  &  1  &  $6.88\begin{array}{l}\scriptstyle +15.81\\\scriptstyle -5.69\end{array}$    &  9.30  &  $34.05\begin{array}{l}\scriptstyle +78.33\\\scriptstyle -28.16\end{array}$  \\
\hline
&&&& \\
 \multicolumn{5}{l}{$^{a}$ Differential number counts} \\ 
 \multicolumn{5}{l}{$^{b}$ Cumulative number counts} \\
\end{tabular}
\label{cn8}
\end{center}
\end{table}

We plot our true 450\,$\mu$m differential number counts (black squares) and true number counts curves from our Monte Carlo simulations (black curves) in Figure \ref{counts}(a), along with the results from {\it Herschel} (\citealt{Oliver:2010p11204}) at 350\,$\mu$m and 500\,$\mu$m (olive and pink squares, respectively). In Figure~\ref{counts}(b), we plot our true 850\,$\mu$m differential number counts (black circles) and true number counts curves from our Monte Carlo simulations (black curves), along with the results of previous SCUBA 850\,$\mu$m lensing cluster surveys given in \citet{Knudsen:2008p3824} (purple filled circles) and of the SHADES SCUBA blank-field survey (\citealt{Coppin:2006p9123}; blue filled circles). We also show the Knudsen et al.\ (2008) 850\,$\mu$m fit to their data (blue curve). In Figures~\ref{counts}(c) and (d), we plot the cumulative number counts at 450\,$\mu$m and 850\,$\mu$m, respectively, which we calculated by summing all the source densities above a given flux.

At 450\,$\mu$m, our bright-end number counts agree with the faint-end counts from {\it Herschel} (Figure \ref{counts}); however, our observations probe a factor of $\sim4$ deeper. Above $\sim30$~mJy, our curves slightly underestimate the {\it Herschel} counts, but our survey area is not large enough to make this result significant. Because of the steeply rising counts as we probe to fainter source fluxes, our 450\,$\mu$m source surface density increases by more than a factor of 10 compared with the results from {\it Herschel}.

With our 450\,$\mu$m true number counts curve, the amount of 450\,$\mu$m EBL we directly resolve is $\sim$ 66.4~Jy/deg$^2$ above 4.5~mJy. This is $\sim 47-61$\% of the total 450\,$\mu$m EBL, depending on the adopted model (\citealt{Puget:1996p2082, Fixsen:1998p2076}). We expect to resolve 100\% of the 450\,$\mu$m EBL at $\sim$ 1 mJy based on our counts model, although we note that since our model curve does not converge, the curve must turn over to a shallower slope in the fainter flux regime. Recently, the closest measurements to our results are from the deep 500\,$\mu$m surveys using {\it Herschel}. Again, the confusion limit caused by {\it Herschel's\/} small aperture size (large beam size) hampers its capability to resolve the EBL (6\%), and the direct counts from {\it Herschel\/} are a factor of $\sim$4 shallower than the present observations (Figure~\ref{counts}). 

A P(D) analysis using pixel flux distributions was used to probe deeper in the {\it Herschel} maps; however, the results were not well constrained at the fainter end beyond the characteristic flux (\citealt{Glenn:2010kx}). \citet{Bethermin:2012yq} performed a stacking analysis on a deep {\it Herschel} 500\,$\mu$m map using a sample of 24\,$\mu$m sources. They found they were able to resolve 55\% of the EBL down to 2~mJy. However, a stacking analysis can only provide information on the selected sample as a whole and does not reveal the true flux distribution of the individual sources. Our observations directly resolve the contributing sources, clearly reveal the characteristic flux, and provide an unbiased sample of sources for further study.

Moreover, we estimate that without lensing, the deboosted counts can be probed to $\sim$8\,mJy at 450\,$\mu$m, corresponding to $\sim$30$-$40\% of the 450 EBL. An extremely deep SCUBA-2 450\,$\mu$m blank-field observation can resolve $\sim$70$-$90\% of the EBL before hitting the confusion limit (estimated at $\sim$2\,mJy given the source density reaching 1 per 30 beam area) based on our current counts model. However, based on the current sensitivity of SCUBA-2, taking into account flux boosting, to directly detect sources with 2\,mJy 450 micron intrinsic fluxes at 4 sigma, one needs more than 100 hours of observing time, which is not practical. Thus to fully resolve 450 micron EBL using SCUBA-2, we will need to observe the lensing fields.

At 850\,$\mu$m, our number counts agree nicely with previous SCUBA results in both blank-field and cluster lensed surveys. Based on our true number counts curve, our observations are able to resolve directly $\sim 40-57$\% of the 850\,$\mu$m EBL above 1.1~mJy with the range again corresponding to the uncertainty in the EBL determination. Our 850\,$\mu$m observations are not as sensitive as many of the confusion-limited blank-field SCUBA surveys; however, thanks to the gravitational lensing, we are able to resolve more EBL than what has been reported from the blank-field SCUBA surveys above 2~mJy ($20-30$\%; e.g., \citealt{Barger:1999p6485}).

\section{450-850-radio comparison}

The tight, universal correlation between FIR and radio luminosity among normal galaxies in the local Universe (\citealt{Helou:1985qy, Condon:1992p6652}) has been used to find the counterparts to the 850\,$\mu$m SMGs and to estimate their redshifts (e.g., \citealt{Carilli:1999p6658, Barger:2000p2144}). However, this method has not been tested on 450\,$\mu$m selected sources. With our newly detected 450\,$\mu$m sources and the deep 20~cm map (1\,$\sigma \sim$ 5.7\,$\mu$Jy) of the A370 field (\citealt{Wold:2012lr}), we are ideally equipped to conduct such a study. Here we study the properties of our submillimeter samples in terms of the source matches, the positional uncertainties, and the redshift distribution estimated through the various flux ratios.

Note that the differential number counts are computed by dividing the summed source surface densities within a flux bin by the flux bin width, which means that the boosting effect on the number counts comes from the mean flux boosting factors (red curves in Figure \ref{flxbst}). This is shown in Section 4.3 where we used the mean flux boosting factors to deboost the raw number counts and to compute the true number counts shown in Figure \ref{counts} that agrees with the underlying true number counts models. 

However, to deboost the fluxes of individual sources for characteristic analysis, the median flux ratios (blue curves in Figure \ref{flxbst}) are more representative for the majority of the population because of the non-Gaussian distribution caused by fake sources with extreme output-to-input flux ratios, especially at lower S/N (S/N $<$ 6; Figure \ref{flxbst}). The median flux ratios can be significantly different than the mean flux ratios, a factor of $\sim 2.7$ versus $\sim 1.6$ at $4~\sigma$ at 450\,$\mu$m. Thus, for the rest of this paper, when we refer to the intrinsic submillimeter fluxes for individual sources, they are delensed and deboosted, where the latter has been done using the median flux boosting factors at the source S/N.

\subsection{Source Matches}

Due to the negative {\it K-}correction at 850\,$\mu$m and the positive {\it K-}correction at 20~cm (\citealt{Carilli:1999p6658, Barger:2000p2144, Blain:2002p8120}), 850\,$\mu$m SMGs with radio counterparts are biased against high-redshift sources, and only a fraction of the 850\,$\mu$m SMGs are seen in the radio. It has been found that around 60\% of the 850\,$\mu$m SMGs with fluxes $>$ 4 mJy have radio counterparts in radio images with a 1\,$\sigma$ sensitivity of $\sim$ 6\,$\mu$Jy (\citealt{Barger:2000p2144, Wang:2004p2270, Ivison:2007lr, Biggs:2011uq}). However, because of the FIR-radio correlation, as the radio images reach deeper, there should be a higher fraction of SMGs that have radio counterparts. Indeed, the latest study by Barger et al. (2012) show that very deep 20~cm interferometric observations with a 1\,$\sigma$ sensitivity of $\sim2.5~\mu$Jy can find all SMGs with 850\,$\mu$m fluxes above 3~mJy.

We cross-identified our submillimeter catalogs with the $>5~\sigma$ radio catalog of \citet{Wold:2012lr}, created from a 20~cm survey of A370 with a 1\,$\sigma$ sensitivity of $\sim$ 6\,$\mu$Jy. We found that 8 out of 18 (44\%) of the 850\,$\mu$m sources with intrinsic fluxes $>4$~mJy have robust radio counterparts within the 850\,$\mu$m beam at the $>5~\sigma$ level. 

We may reduce our radio S/N threshold if we have more accurate submillimeter positions. We ran a simulation to estimate the probability of a randomly positioned 450\,$\mu$m beam selecting a $>3~\sigma$ random noise fluctuation on a simulated normal-distributed radio map. We found the probability to be $\sim5$\%. The probability should be even lower if there are counterparts at other wavelengths (e.g., 850\,$\mu$m). Note that \citet{Biggs:2011uq} considered radio sources with 5\% of the so-called corrected Poissonian probability (\citealt{Browne:1978vn}) to be secure counterparts for SMGs.  Using the 450\,$\mu$m data, we identified four more $>3~\sigma$ radio counterparts to the 850\,$\mu$m sources. Thus, in total, 12/18 or 66.7\% of the  850\,$\mu$m sources with intrinsic fluxes $>4$~mJy have radio matches, which is consistent with what has been reported in the literature.

For the 450\,$\mu$m sources, the fraction having secure ($>3~\sigma$) radio counterparts is 85\% (17/20). This high match rate could mean that both the 450\,$\mu$m sources with intrinsic fluxes $>6$~mJy and the 20~cm emission are preferentially tracing lower redshift starbursting galaxies, because both have positive {\it K-}corrections at high redshifts. Meanwhile, 67\% (8/12) of the $>4~\sigma$ 450\,$\mu$m sources are detected at 850\,$\mu$m, while the recovery rate for $>4~\sigma$ 850\,$\mu$m sources at 450\,$\mu$m is slightly lower (13/24; 54\%). 

The moderate match between 450\,$\mu$m and 850\,$\mu$m sample again implies that the SMG population is diverse and different selection methods reveal only parts of the overall population. For example, SMGs detected only at 450\,$\mu$m tend to have a higher dust temperature and/or lower redshift (\citealt{Casey:2012fk}), whereas the 850\,$\mu$m sources without radio or 450\,$\mu$m counterparts provide the most probable candidate high-redshift sources (e.g., \citealt{Riechers:2010p9307, Walter:2012qy}).

\subsection{Positional Uncertainties}
\begin{figure}[h]
 \begin{center}
    \leavevmode
      \includegraphics[scale=0.72]{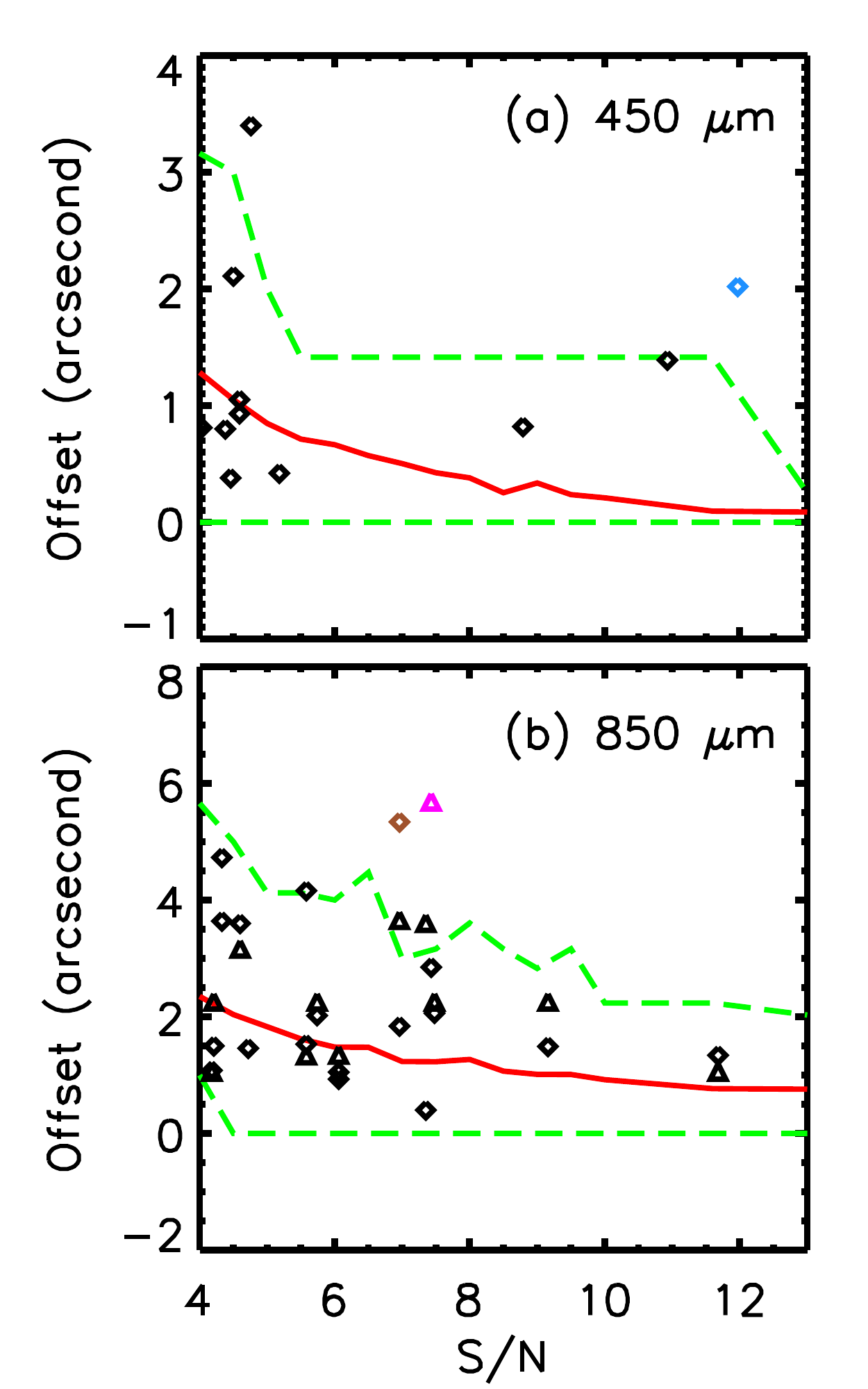}
       \caption{The positional offsets between the output and the brightest input sources within the beam area from our Monte Carlo simulations (red curves). The areas enclosed by the green dashed curves represent the 90\% confidence ranges relative to the mean values. The diamonds are the offsets between our cataloged (a) 450\,$\mu$m sources and (b) 850\,$\mu$m sources relative to the 5\,$\sigma$ radio sources. The triangles in (b) are the offsets between the 450\,$\mu$m and 850\,$\mu$m sources. A370-450.1/A370-850.1, A370-450.18/A370-850.7 and A370-450.9/A370-450.11/A370-850.5 are plotted with the blue diamond in (a), the brown diamond and the magenta triangle in (b), respectively. 
       }
     \label{poserr}
  \end{center}
\end{figure}

In the absence of 450\,$\mu$m observations or submillimeter or millimeter interferometry, radio sources have been widely used to locate counterparts to the SMGs detected at 850\,$\mu$m by single-dish telescopes (e.g., \citealt{Barger:2000p2144, Smail:2002p6793, Chapman:2005p5778, Pope:2005uq}). This method has been shown to be appropriate for many of the sources using high-resolution submillimeter interferometric observations, though in some cases there may be multiple counterparts rather than a single source (\citealt{Younger:2007p6982, Younger:2009p9502, Wang:2011p9293, Barger:2012lp, Karim:2012lr}). In contrast to following up each source individually with interferometers,  we can statistically examine the method by comparing the positional offsets between radio and submillimeter sources from real observations and the offsets obtained from Monte Carlo simulations. In theory, if the radio emission traces the submillimeter emission, then the observational results should agree statistically with the simulations. 

The details of the simulations are described in Section~4.2. We note that in the simulations, at both wavelengths we treated the brightest input source as the counterpart, and we gridded the maps to 1$'' \times 1''$ in order to match with the real submillimeter data maps. We plot our results in Figure~\ref{poserr} at (a) 450\,$\mu$m and (b) 850\,$\mu$m. The observational offsets between the submillimeter sources and their secure radio counterpart candidates are in diamonds in both panels, and the simulated mean offsets are denoted by red curves with the 90\% confidence range enclosed by green dashed curves. The observational results generally agree with our simulations, except for two sources, A370-450.1/A370-850.1 (blue diamond in Figure \ref{poserr}(a)) and A370-450.18/A370-850.7 (brown diamond in Figure \ref{poserr}(b)). A detailed analysis of these two sources is given in the later sections and the Appendix, in which we conclude that those two sources are likely to be complex systems with multiple sources contributing emission at different wavelengths. Taking A370-450.1/A370-850.1 as an example, recent studies of this source have revealed its complex nature, including multiple sources having different characteristics being located within $1-2''$ of one another (more details are given on this source in the Appendix) and evidence for a significant offset between the dust and radio emission (\citealt{Genzel:2003kx, Ivison:2010yq}). Our statistical diagnosis serves as a simple test as to whether there is a common origin for the emission seen in the different wavebands.

We also plot the offsets between the 450\,$\mu$m and 850\,$\mu$m sources on Figure~\ref{poserr}(b) (black triangles). Again, we find good statistical agreement between the observational offsets and the offsets that we obtain from the simulations, which supports the idea that the 450\,$\mu$m sources are a good positional tracer for the 850\,$\mu$m sources. Although the 450\,$\mu$m positions are less accurate than the radio positions, the 450\,$\mu$m emission has the advantage of directly tracing the thermal dust emission, just as the 850\,$\mu$m emission does. 

Similar to the situation for A370-450.1 mentioned above, there is a complex system in the submillimeter, A370-450.9/A370-450.11/A370-850.5, that lies off the simulated curve in Figure~\ref{poserr}(b) (magenta triangle). This could be another example of a multiple, where different sources contribute emission at different wavelengths, and sources selected at different wavelengths may have different origins (\citealt{Wang:2011p9293};  Barger et al. 2012). Note that in this plot we show the offset between A370-450.11 and A370-850.5. The offset between A370-450.9 and A370-850.5 is even larger.

In general, there is a slight upward bias for the real data relative to the simulated mean curve, which could be caused by the fact that some of the counterparts may not be correct. As we show in subsequent sections, some of the sources have multiple radio counterparts within the beam area, which is an indication of a complex nature. Including incorrect counterparts leads to a large dispersion and, in our case, an upward bias.

\subsection{Millimetric Redshifts}

\begin{figure}[h]
 \begin{center}
    \leavevmode
      \includegraphics[scale=0.32]{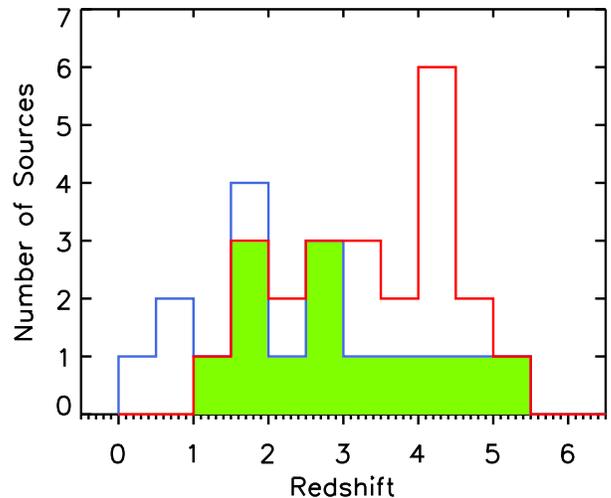}
       \caption{Histograms of the source redshifts from Table~\ref{z}. The redshift distribution for the sources detected at both 450\,$\mu$m and 850\,$\mu$m (green shading) has a median $z\sim2.5$. The red histogram outlines the redshift distribution for the 850\,$\mu$m sources only, while the blue histogram outlines the redshift distribution for the 450\,$\mu$m sources only. The median redshift for all the 450 (850)\,$\mu$m sources is $z\sim2.3$ ($z\sim3.4$). A370-450.13 is not included in this plot, since in this case the redshifts estimated from the different flux ratios diverge.}
     \label{zhisto}
  \end{center}
\end{figure}

We estimated the redshifts of each source using the 850\,$\mu$m-to-450\,$\mu$m and submillimeter-to-20~cm flux ratios assuming a modified blackbody SED template with $\beta = 1$ (dust emissivity) and $T_d = 47$~K (dust temperature) from the local starbursting galaxy Arp~220 (\citealt{Klaas:1997kx, Barger:2000p2144}). In Table \ref{z}, we show the redshifts of each individual source estimated using three different flux ratios. We show dots when there is not a measurement for both fluxes (e.g., $z_{450/radio}$ on A370-850.11) or when there are multiple counterpart candidates (e.g., A370-450.17/850.4 with its multiple radio counterparts). In many cases, the redshift estimates using different flux ratios, in particular 850\,$\mu$m-to-450\,$\mu$m and 850\,$\mu$m-to-20~cm, agree with one other to within the uncertainties. A noticeable disagreement is seen for SMGs not detected at 850\,$\mu$m (e.g., A370-450.6, 450.8, 450.10, 450.12). Here the flux ratios between 850\,$\mu$m and 450\,$\mu$m indicate a low redshift, while the flux ratios between 450\,$\mu$m-and 20~cm indicate higher redshifts. Given that 450\,$\mu$m data suffer from a positive {\it K-}correction, much as 20~cm data do, such sources are more likely to be at low redshifts. The fact they have abnormally high 450\,$\mu$m fluxes probably indicates that they have higher dust temperatures than what we assumed in the SED template. Thus, the 850\,$\mu$m fluxes lie below the detection limit.  

In the final column of Table~\ref{z}, we give our adopted redshift for each source that had good constraints on the redshift estimation. Redshifts with three digits after the decimal point represent the spectroscopic measurements. We primarily adopted the redshifts estimated through the 850\,$\mu$m-to-20~cm flux ratio, since it has recently been shown to be a reasonable redshift estimator with good accuracies (Barger et al.\ 2012). We adopted the redshifts estimated from the 850\,$\mu$m-to-450\,$\mu$m flux ratio whenever the 850\,$\mu$m-to-20~cm flux ratio was unavailable. 

We plot histograms of our redshifts in Figure~\ref{zhisto}. Sources detected at both 450\,$\mu$m and 850\,$\mu$m (green shading) have a median redshift of $z\sim2.5$. The redshift distribution of the sources detected only at 450 (850)\,$\mu$m is denoted by blue (red) lines. We adopted the upper limits of the redshifts estimated using the 850\,$\mu$m-to-450\,$\mu$m flux ratio for the sources detected only at 450\,$\mu$m. For 850\,$\mu$m sources without good redshift constraints, we adopted the lower limits estimated using the 850\,$\mu$m-to-20~cm flux ratios. In the final column of Table~\ref{z}, we show these redshift limits with parenthesis. All the 850\,$\mu$m sources have a median redshift of $z\sim3.4$, while all the 450\,$\mu$m sources have a median redshift of $z\sim2.3$. These results generally agree with the picture that 450\,$\mu$m sources preferentially trace lower redshift dusty sources, while 850\,$\mu$m sources tend towards higher redshifts. We do, however, caution that because there are many issues about fitting the FIR SED template for SMGs such as the degeneracy between source redshift, dust temperature and dust emissivity (\citealt{Casey:2012rt}), our results are first order estimations and should be used carefully.

\begin{table}
\caption{Redshifts}
\centering
\scalebox{0.82}{
\begin{tabular}{lllllr}
\hline
 ID$_{450}$      &  ID$_{850}$      &  z$_{850/450}$                                                                  &  z$_{450/radio}$                                                                  &  z$_{850/radio}$                                                                  &  z$_{adopted}^a$ \\
\hline
 A370-450.1   &  A370-850.1   &  \ldots  &  \ldots                                                                   &  \ldots                                                                   &  2.800   \\
 A370-450.2   &  A370-850.3   &  $0.00\begin{array}{l}\scriptstyle +0.76\\\scriptstyle -0.00\end{array}$  &  $0.97\begin{array}{l}\scriptstyle +0.07\\\scriptstyle -0.07\end{array}$  &  $0.84\begin{array}{l}\scriptstyle +0.07\\\scriptstyle -0.07\end{array}$  &  1.060   \\
 A370-450.3   &  A370-850.2   &  $2.08\begin{array}{l}\scriptstyle +0.81\\\scriptstyle -0.84\end{array}$  &  $1.85\begin{array}{l}\scriptstyle +0.19\\\scriptstyle -0.18\end{array}$  &  $1.90\begin{array}{l}\scriptstyle +0.13\\\scriptstyle -0.12\end{array}$  &  $1.90\begin{array}{l}\scriptstyle +0.13\\\scriptstyle -0.12\end{array}$  \\
 A370-450.4   &  A370-850.18  &  $0.00\begin{array}{l}\scriptstyle +1.73\\\scriptstyle -0.00\end{array}$  &  $1.43\begin{array}{l}\scriptstyle +0.22\\\scriptstyle -0.22\end{array}$  &  $1.18\begin{array}{l}\scriptstyle +0.18\\\scriptstyle -0.20\end{array}$  &  1.519   \\
 A370-450.5   &  A370-850.6   &  $3.48\begin{array}{l}\scriptstyle +1.30\\\scriptstyle -1.31\end{array}$  &  $5.95\begin{array}{l}\scriptstyle +0.00\\\scriptstyle -2.30\end{array}$  &  $4.47\begin{array}{l}\scriptstyle +1.15\\\scriptstyle -0.76\end{array}$  &  $4.47\begin{array}{l}\scriptstyle +1.15\\\scriptstyle -0.76\end{array}$  \\
 A370-450.6   &  \ldots          &  $<$0.01                                                                &  $5.95\begin{array}{l}\scriptstyle +0.00\\\scriptstyle -0.98\end{array}$  &  \ldots                                                                   & (0.01)  \\
 A370-450.7   &  A370-850.8   &  $2.52\begin{array}{l}\scriptstyle +1.50\\\scriptstyle -1.61\end{array}$  &  \ldots  &  \ldots  &  $2.52\begin{array}{l}\scriptstyle +1.50\\\scriptstyle -1.61\end{array}$  \\
 A370-450.8   &  \ldots          &  $<$0.80                                                                &  $1.74\begin{array}{l}\scriptstyle +0.36\\\scriptstyle -0.33\end{array}$  &  \ldots                                                                   &  (0.80)  \\
 A370-450.9   &  A370-850.5   &  \ldots                                                                &  \ldots                                                                   &  \ldots                                                                   &  \ldots  \\
 A370-450.10  &  \ldots          &  $<$0.94                                                                &  $3.08\begin{array}{l}\scriptstyle +1.48\\\scriptstyle -0.79\end{array}$  &  \ldots                                                                   &  (0.94)  \\
 A370-450.11  &  A370-850.5   &  \ldots                                                                &  \ldots                                                                   &  \ldots                                                                   &  \ldots  \\
 A370-450.12  &  \ldots          &  $<$1.80                                                                &  $3.67\begin{array}{l}\scriptstyle +2.28\\\scriptstyle -1.13\end{array}$  &  \ldots                                                                   &  (1.80)  \\
 A370-450.13  &  A370-850.26  &  $0.66\begin{array}{l}\scriptstyle +2.47\\\scriptstyle -0.66\end{array}$  &  $>$5.95                                                                &  $>$3.22                                                                &  \ldots  \\
 A370-450.14  &  A370-850.17  &  $1.52\begin{array}{l}\scriptstyle +2.07\\\scriptstyle -1.52\end{array}$  &  $2.99\begin{array}{l}\scriptstyle +1.55\\\scriptstyle -0.83\end{array}$  &  $2.53\begin{array}{l}\scriptstyle +0.58\\\scriptstyle -0.49\end{array}$  &  $2.53\begin{array}{l}\scriptstyle +0.58\\\scriptstyle -0.49\end{array}$  \\
 A370-450.15  &  A370-850.9   &  $3.31\begin{array}{l}\scriptstyle +1.68\\\scriptstyle -1.70\end{array}$  &  $1.95\begin{array}{l}\scriptstyle +0.48\\\scriptstyle -0.43\end{array}$  &  $2.29\begin{array}{l}\scriptstyle +0.31\\\scriptstyle -0.30\end{array}$  &  $2.29\begin{array}{l}\scriptstyle +0.31\\\scriptstyle -0.30\end{array}$  \\
 A370-450.16  &  A370-850.10  &  $3.63\begin{array}{l}\scriptstyle +1.70\\\scriptstyle -1.70\end{array}$  &  \ldots                                                                   &  \ldots                                                                   &  $3.63\begin{array}{l}\scriptstyle +1.70\\\scriptstyle -1.70\end{array}$  \\
 A370-450.17  &  A370-850.4   &  $4.61\begin{array}{l}\scriptstyle +1.60\\\scriptstyle -1.45\end{array}$  &  \ldots                                                                   &  \ldots   &  $4.61\begin{array}{l}\scriptstyle +1.60\\\scriptstyle -1.45\end{array}$  \\
 A370-450.18  &  A370-850.7   &  $5.01\begin{array}{l}\scriptstyle +1.71\\\scriptstyle -1.52\end{array}$  &  \ldots                                                                   &  \ldots  &  $5.01\begin{array}{l}\scriptstyle +1.71\\\scriptstyle -1.52\end{array}$  \\
 A370-450.19  &  A370-850.25  &  $2.28\begin{array}{l}\scriptstyle +2.28\\\scriptstyle -2.28\end{array}$  &  $1.42\begin{array}{l}\scriptstyle +0.34\\\scriptstyle -0.35\end{array}$  &  $1.58\begin{array}{l}\scriptstyle +0.25\\\scriptstyle -0.27\end{array}$  &  $1.58\begin{array}{l}\scriptstyle +0.25\\\scriptstyle -0.27\end{array}$  \\
 A370-450.20  &  A370-850.15  &  $2.80\begin{array}{l}\scriptstyle +2.18\\\scriptstyle -2.38\end{array}$  &  $3.04\begin{array}{l}\scriptstyle +1.87\\\scriptstyle -0.94\end{array}$  &  $3.08\begin{array}{l}\scriptstyle +0.65\\\scriptstyle -0.55\end{array}$  &  $3.08\begin{array}{l}\scriptstyle +0.65\\\scriptstyle -0.55\end{array}$  \\
 \ldots          &  A370-850.11  &  $>$2.93                                                                &  \ldots                                                                   &  $>$4.29                                                                &  (4.29)  \\
 \ldots          &  A370-850.12  &  $>$2.14                                                                &  \ldots                                                                   &  $>$4.25                                                                &  (4.25)  \\
 \ldots          &  A370-850.13  &  $>$1.98                                                                &  \ldots                                                                   &  $>$4.71                                                                &  (4.71)  \\
 \ldots          &  A370-850.14  &  $>$1.99                                                                &  \ldots                                                                   &  $2.34\begin{array}{l}\scriptstyle +0.39\\\scriptstyle -0.37\end{array}$  &  $2.34\begin{array}{l}\scriptstyle +0.39\\\scriptstyle -0.37\end{array}$  \\
 \ldots          &  A370-850.16  &  $>$1.78                                                                &  \ldots                                                                   &  \ldots                                                                &  \ldots  \\
 \ldots          &  A370-850.19  &  $>$0.81                                                                &  \ldots                                                                   &  $>$4.12                                                                &  (4.12)  \\
 \ldots          &  A370-850.20  &  $>$0.00                                                                &  \ldots                                                                   &  $>$4.49                                                                &  (4.49)  \\
 \ldots          &  A370-850.21  &  $>$1.50                                                                &  \ldots                                                                   &  $>$4.12                                                                &  (4.12)  \\
 \ldots          &  A370-850.22  &  $>$1.31                                                                &  \ldots                                                                   &  $>$3.61                                                                &  (3.61)  \\
 \ldots          &  A370-850.23  &  $>$1.23                                                                &  \ldots                                                                   &  $3.08\begin{array}{l}\scriptstyle +0.79\\\scriptstyle -0.63\end{array}$  &  $3.08\begin{array}{l}\scriptstyle +0.79\\\scriptstyle -0.63\end{array}$  \\
 \ldots          &  A370-850.24  &  $>$0.48                                                                &  \ldots                                                                   &  $>$3.41                                                                &  (3.41)  \\
\hline
&&&&& \\
 \multicolumn{6}{l}{$^{a}$Values with three digits after the decimal point are spectroscopic redshifts.} \\ 
\end{tabular}
}
\label{z}
\end{table}

\begin{figure}[h]
 \begin{center}
    \leavevmode
      \includegraphics[scale=0.35]{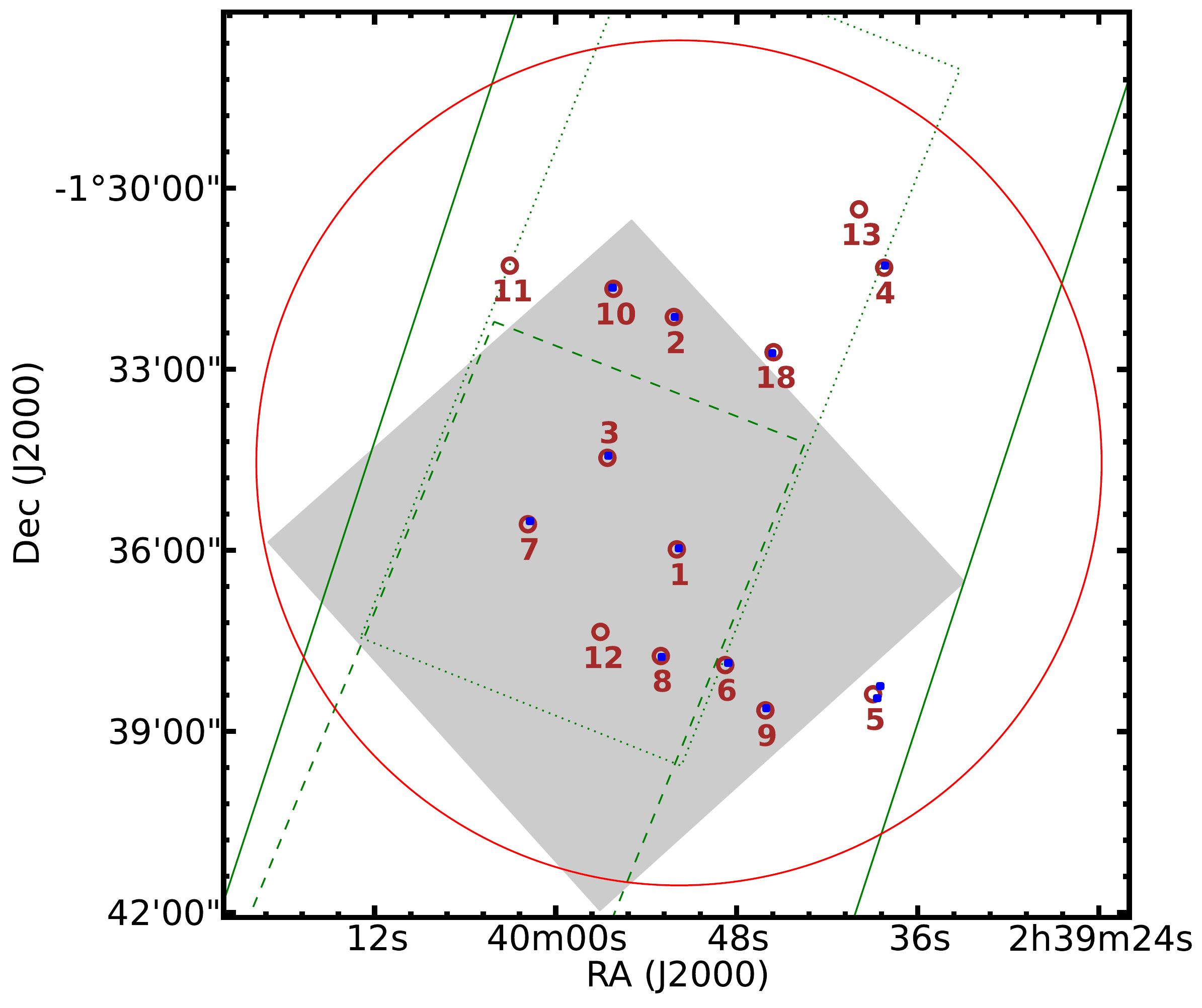}
       \caption{The schematic sky coverage of the A370 field. The approximate SCUBA-2 coverage is denoted by the red circle with a radius of 7$'$. The {\it Spitzer\/} 3.6\,$\mu$m and 4.5\,$\mu$m coverage is marked with the green solid lines, and the 5.8\,$\mu$m and 8.0\,$\mu$m coverages are marked, respectively, with the green dashed and dotted lines. The blue dots (red circles) represent the locations of our 450 (850)\,$\mu$m sources in our SCUBA-2 robust sample. The numbers given are the identifications of the 850\,$\mu$m sources. The gray area shows the X-ray coverage obtained by {\it Chandra}. 
       }
     \label{coverage}
  \end{center}
\end{figure}

\begin{figure*}[t]
 \begin{center}
    \leavevmode
      \includegraphics[scale=1.18]{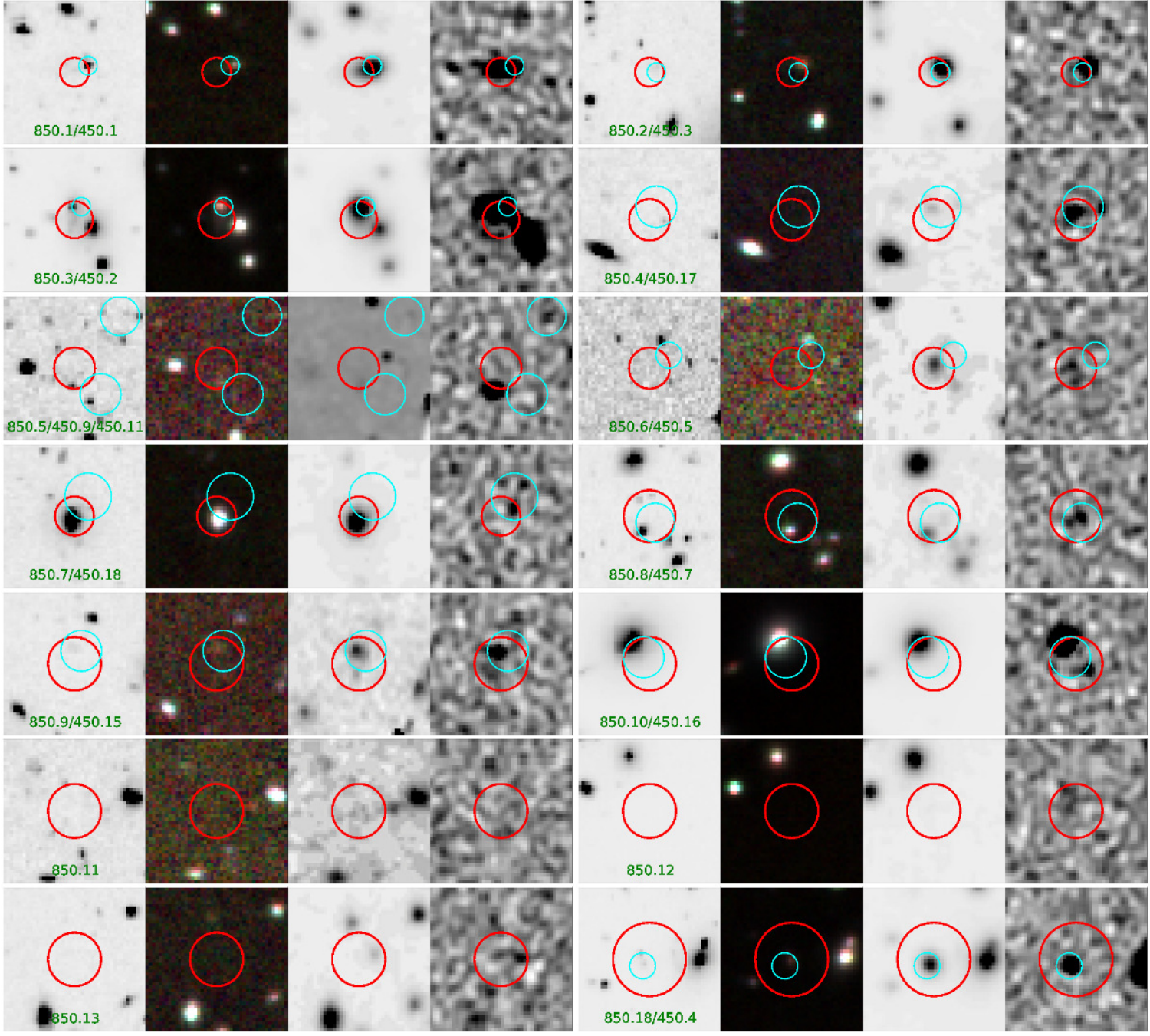}
       \caption{Postage stamp images of the 5\,$\sigma$ submillimeter sources in our SCUBA-2 robust sample. The images from left to right in each panel are an inverse gray scale $z$-band image, a false color near-Infrared ($J, H, K$) image, an inverse gray scale 4.5\,$\mu$m image, and an inverse gray scale 20~cm image. The images are plotted with a dynamical range from 1\% to 99\% using a linear scale. The angular size of each image is 22$'' \times 22''$. Cyan and red circles represent, respectively, the positional uncertainties with 95\% confidence at 850\,$\mu$m and 450\,$\mu$m (see Section~5.1). 
       }
     \label{comb_1}
  \end{center}
\end{figure*}

\section{Counterpart Identifications and Discussion}

\begin{table*}
\caption{NIR, MIR, and Radio photometry}
\centering
\scalebox{0.98}{
\begin{tabular}{lllrrrrrrlrrl}
\hline
\hline
 Source ID       &  R.A.$^a$       &  DEC.$^a$         &  offset$^b$  &     $J$  &     $H$  &     $K_s$  &  3.6\,$\mu$m  &  4.5\,$\mu$m  &  5.8\,$\mu$m  &  8.0\,$\mu$m  &  20 cm  &  z  \\
			& (J2000.0) 	& (J2000.0)   & ($''$)	&	(mag)	&	(mag	) &	(mag	) &	(mag)	&	(mag) 	&	(mag) 	&	(mag)	& ($\mu$Jy) & \\
\hline
 850.1/450.1     &  2 39 52.0  &  -1 35 58.5  &  2.0  &   21.8  &   21.4  &   21.2  &    20.4  &   20.0  &  19.4    &    18.9  &  164.5  &  2.8000  \\
 850.2/450.3     &  2 39 52.1  &  -1 32 07.2  &  0.8  &  21.8  &   21.0  &   20.2  &    19.2  &   18.9  &  \ldots  &    19.1  &  150.4  &  \ldots  \\
 850.3/450.2     &  2 39 56.6  &  -1 34 26.6  &  1.3  &  19.9  &   19.7  &   19.2  &    18.4  &   18.4  &  18.3    &    18.3  &  245.5  &  1.0600  \\
 850.4/450.17a   &  2 39 38.3  &  -1 31 17.4  &  2.1  & $>$23.5  &  $>$23.0  &  $>$23.0  &    22.8  &   22.1  &  \ldots  &  \ldots  &   61.7  &  \ldots  \\
 850.4/450.17b   &  2 39 38.0  &  -1 31 16.6  &  2.4  &  $>$23.5  &  $>$23.0  &  $>$23.0  &    23.4  &   22.7  &  \ldots  &  \ldots  &   25.8  &  \ldots  \\
 850.5/450.11a   &  2 39 39.0  &  -1 38 25.8  &  2.9  &  23.2  &   22.6  &   21.9  &  \ldots  &   22.0  &  \ldots  &  \ldots  &  275.0  &  3.8219  \\
 850.5/450.11b   &  2 39 38.8  &  -1 38 28.1  &  2.0  &   23.0  &  $>$23.0  &  $>$23.0  &   $>$23.3  &  $>$23.3  &  $>$23.0   &   $>$22.6  &  $<$16.7  &  \ldots  \\
 850.5/450.9     &  2 39 38.5  &  -1 38 14.6  &  0.4  &  $>$23.5  &  $>$23.0  &  $>$23.0  &  \ldots  &   22.4  &  \ldots  &  \ldots  &   18.4  &  \ldots  \\
 850.6/450.5     &  2 39 48.8  &  -1 37 53.6  &  3.9  &   $>$23.6  &  $>$23.1  &  $>$23.1  &    21.9  &   21.3  &  \ldots  &  \ldots  &   18.3  &  \ldots  \\
 850.7/450.18a   &  2 40 01.8  &  -1 35 34.8  &  4.0  &  18.9  &   18.8  &   18.6  &    19.2  &   19.4  &  19.7    &    20.2  &  $<$14.4  &  \ldots  \\
 850.7/450.18b   &  2 40 01.7  &  -1 35 33.2  &  2.2  & $>$23.7  &  $>$23.2  &  $>$23.2  &   $>$23.5  &  $>$23.2  &  $>$23.2   &   $>$22.8  &   23.3  &  \ldots  \\
 850.7/450.18c   &  2 40 01.6  &  -1 35 29.7  &  2.0  &  $>$23.7  &  $>$23.2  &  $>$23.2  &   $>$23.5  &  $>$23.2  &  $>$23.2   &   $>$22.8  &   17.9  &  \ldots  \\
 850.8/450.7a    &  2 39 53.1  &  -1 37 47.5  &  2.5  &   20.2  &   20.3  &   20.1  &    20.7  &   20.9  &  20.9    &    22.1  &  $<$14.8  &  \ldots  \\
 850.8/450.7b    &  2 39 52.9  &  -1 37 45.3  &  1.3  & $>$23.7  &  $>$23.2  &  $>$23.2  &    22.5*  &   21.5*  &  22.3*    &    21.4*  &   23.8  &  \ldots  \\
 850.8/450.7c    &  2 39 53.0  &  -1 37 45.9  &  0.5  & $>$23.7  &  $>$23.2  &  $>$23.2  &    22.5*  &   21.5*  &  22.3*    &    21.4*  &   15.2  &  \ldots  \\
 850.9/450.15     &  2 39 46.1  &  -1 38 37.0  &  1.1  & 23.3  &   22.2  &   22.2  &    21.2  &   20.6  &  \ldots  &  \ldots  &   67.6  &  \ldots  \\
 850.10/450.16a  &  2 39 56.3  &  -1 31 36.7  &  2.5  &  16.0  &   16.1  &   16.3  &    16.4  &   16.6  &  \ldots  &    14.3  &  938.1  &  0.0286  \\
 850.10/450.16b  &  2 39 56.1  &  -1 31 41.1  &  2.9  &  $>$23.7  &  $>$23.2  &  $>$23.2  &   $>$23.5  &  $>$23.2  &  \ldots  &   $>$22.8  &   50.0  &  \ldots  \\
 850.13          &  2 39 39.8  &  -1 30 20.2  &  1.7  & $>$23.5  &  $>$23.0  &  $>$23.0  &    22.6  &   21.7  &  \ldots  &    21.2  &   23.1  &  \ldots  \\
 850.18/450.4    &  2 39 45.6  &  -1 32 43.6  &  0.6  & 20.9  &   20.7  &   20.2  &    19.5  &   19.2  &  \ldots  &    18.9  &  144.3  &  1.5190  \\
\hline
&&&&&&&&&&&& \\
\multicolumn{13}{l}{$^{a}$Positions are extracted from the radio counterpart, if present, and otherwise from the optical counterpart. For}\\
\multicolumn{13}{l}{     sources without any radio or optical counterparts, we used the positions from the images with the best resolution.} \\
\multicolumn{13}{l}{$^{b}$Offsets are calculated with the 450\,$\mu$m position as the reference, except for 850.13.}\\
\multicolumn{13}{l}{$^*$Source blended} 

\end{tabular}
}
\label{cntrprt}
\end{table*}

 \begin{figure}[t]
\begin{center}
   \leavevmode
      \includegraphics[scale=0.65]{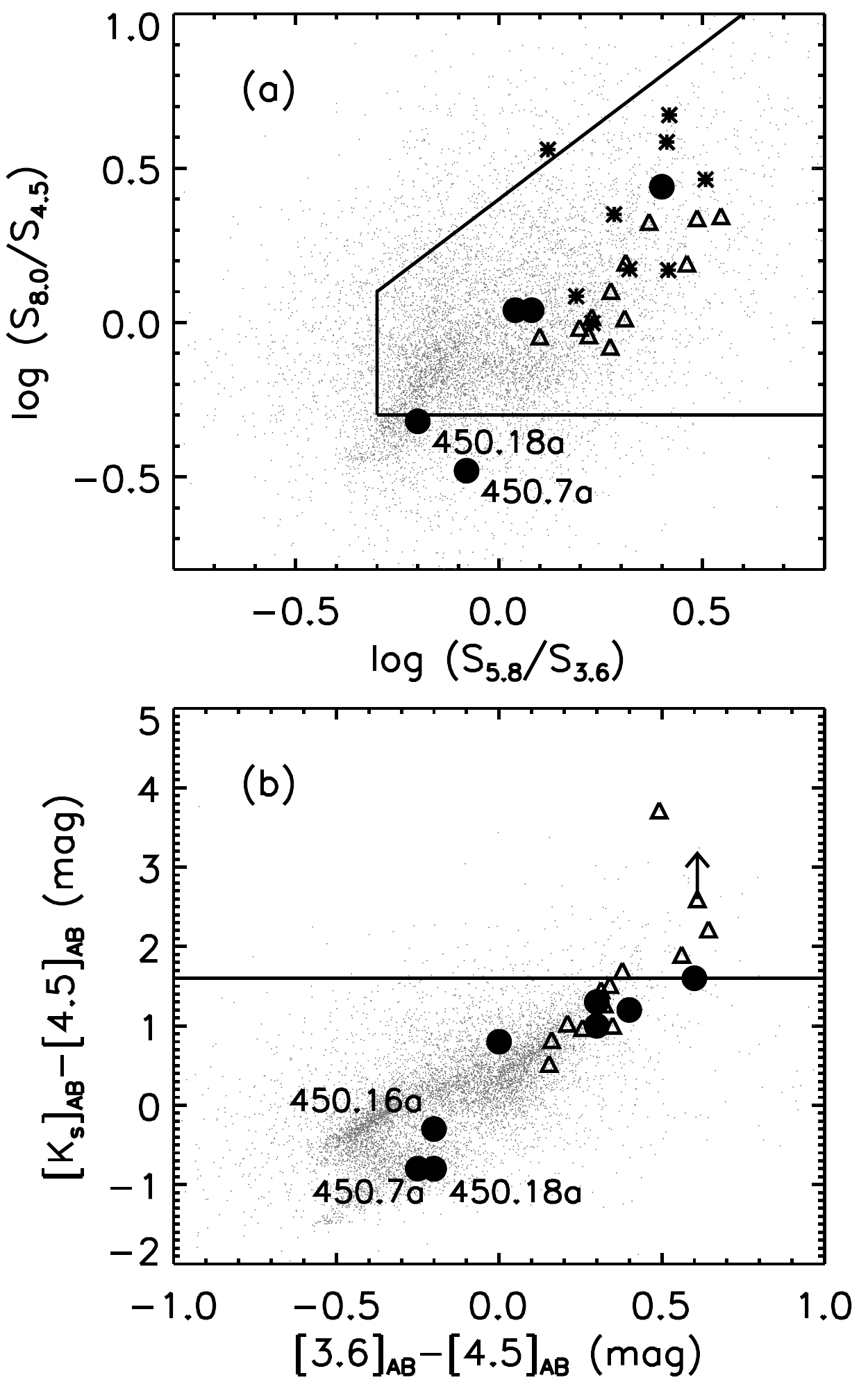}
       \caption{IR color-color diagram with filled circles representing our SCUBA-2 robust sample sources with measurements in all four passbands and triangles showing the GOODS-N SMG sample observed with the SMA by Barger et al.\ (2012). (a) 8.0\,$\mu$m/4.5\,$\mu$m vs. 5.8\,$\mu$m/3.6\,$\mu$m color-color diagram. Another SMG sample obtained with the AzTEC/COSMOS survey at 890\,$\mu$m and observed with the SMA is denoted by asterisks (Younger et al.\ 2007, 2009). The area enclosed by the solid lines is the color space proposed by Yun et al.\ (2008) for selecting SMGs. (b) $K_s$--4.5\,$\mu$m vs. 3.6\,$\mu$m--4.5\,$\mu$m color-color diagram. The field galaxies selected in a deep $K_s$ survey of the GOODS-N field by \citet{Wang:2012ys} are plotted in gray dots, and the area above the horizontal line shows their KIEROs selection ($K_s-4.5$\,$\mu$m (AB) $>$ 1.6). Two sources (450.7a and 450.18a) have the sameflux ratios, so we have moved the data point of 450.7a 0.05~mag to the left for clarity. }
     \label{ir_cc}
  \end{center}
\end{figure}

In this Section, we restrict our sample to the 14 sources that are detected at $>5~\sigma$ in at least one submillimeter waveband for their robust detections and better constraints on the astrometry (Section~5.2). We refer to these 14 sources as our SCUBA-2 robust sample. We note that most part of our 450\,$\mu$m map has a sensitivity lower than the the {\it Herschel} confusion limit (1\,$\sigma\sim$ 6.8 mJy; \citealt{Nguyen:2010kx}) at similar wavelengths, and even at 5\,$\sigma$ level, majority (10 out of 14) of our SCUBA-2 robust sample would not be found by {\it Herschel} at 4\,$\sigma$.

Except for the {\it Spitzer\/} and {\it Chandra\/} data, the data at other wavelengths fully cover the SCUBA-2 surveyed area. We show the schematic {\it Spitzer\/} and {\it Chandra\/} coverage in Figure \ref{coverage}, where the 450\,$\mu$m and 850\,$\mu$m sources in the SCUBA-2 robust sample are marked with red and blue symbols, and the identification numbers of the 850\,$\mu$m sources are given. The {\it Spitzer\/} coverage at various wavelengths is denoted by the green lines, while the gray area shows the area observed by {\it Chandra}. 

In Figure \ref{comb_1}, we give postage stamp images for each of the 14 sources, including an inverse gray scale $z$-band image, a false color $J, H, K_s$ image, an inverse gray scale 4.5\,$\mu$m image, and an inverse gray scale 20~cm image. We use cyan and red circles to represent the positional error of the 450\,$\mu$m and 850\,$\mu$m sources with 95\% confidence. 

In Table \ref{cntrprt}, we give the delensed photometry of each candidate counterpart from the NIR to the radio. We label these candidate counterparts with the submillimeter source identification followed by a small ``a'', ``b'', or ``c'', if there are multiple candidates. We use dots in the table when the submillimeter source lies outside the coverage of a particular wavelength image. We discuss the individual sources in the Appendix. Note that because there are no sources in our SCUBA-2 robust sample that are only detected at 450\,$\mu$m, when referring to individual sources hereafter, we will use the 450\,$\mu$m identifications for sources detected at both 450\,$\mu$m and 850\,$\mu$m, and we will use the 850\,$\mu$m identifications for sources only detected at 850\,$\mu$m.

\subsection{X-ray Candidate Counterparts}

While a number of studies have shown that most of the FIR luminosity from SMGs comes from star formation, the issue of what fraction of SMGs contain an AGN is far from settled. Results have ranged from as many as 75\% (\citealt{Alexander:2005p6453}; radio-detected SMGs) down to $\sim$10--30\% (\citealt{Georgantopoulos:2011mz}; {\em Spitzer}-detected SMGs). Existing {\it Chandra\/} observations of the A370 field partially cover our SCUBA-2 survey area and include 9 850\,$\mu$m sources and 8 450\,$\mu$m sources from the SCUBA-2 robust sample. We cross-correlated the source catalogs in \citet{Barger:2001uq} and found 2 counterparts in hard X-rays (A370-450.1, A370-450.2) and 1 in soft X-rays (A370-450.16). This gives a fraction of 3/9 robust 850\,$\mu$m sources containing an AGN ($\sim$33\%), which is similar to what Georgantopoulos et al.\ (2011) found but much lower than the results of Alexander et al.\ (2005). Note that both of these studies from the literature used much deeper X-ray maps with exposure times of 1~Ms and 2~Ms, respectively. Thus, it is not surprising that we find a lower fraction of submillimeter sources containing an AGN. For our robust 450\,$\mu$m sources, the fraction is 3/8 ($\sim$38\%).

 \begin{figure}[t]
\begin{center}
   \leavevmode
      \includegraphics[scale=0.65]{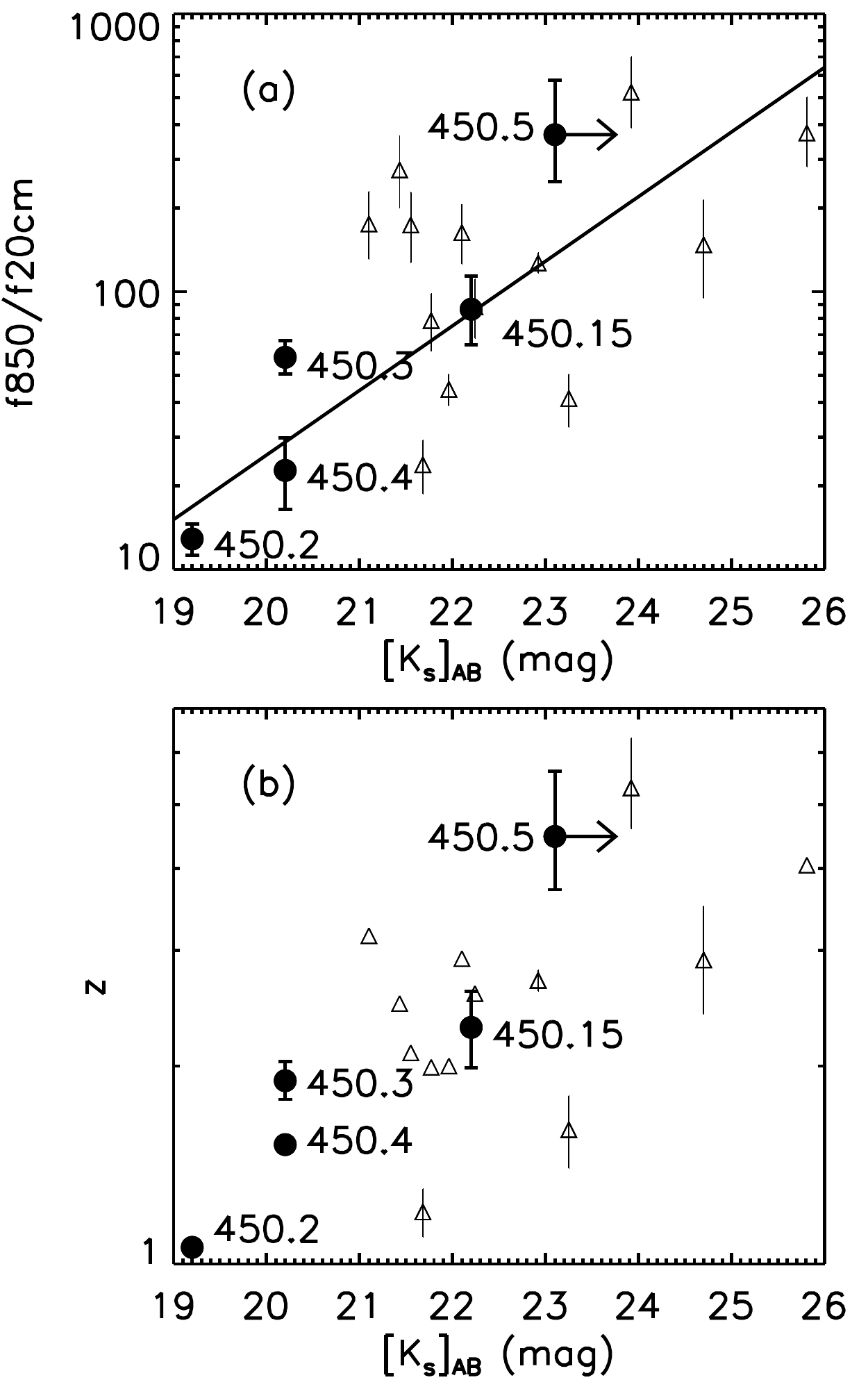}
       \caption{(a) 850\,$\mu$m/20~cm--$K_s$ color-magnitude diagram. Our sample sources are shown in filled circles with their IDs. The linear solid curve represents the minimum $\chi$--sqruare fit to our data points. Results from submillimeter interferometric follow-up on GOODS-N SMGs are plotted in triangles (Barger et al. 2012). (b) Redshift--$K_s$ diagram using the same sample sources as described in (a). Millimetric redshifts are adopted with error bars whenever the spectroscopic redshifts are not available.}
     \label{nir_k82}
  \end{center}
\end{figure}

\subsection{IR Candidate Counterparts}

The IR properties of the 850\,$\mu$m SMGs have been extensively studied from the NIR to {\it Spitzer} 3.6\,$\mu$m to 24\,$\mu$m (\citealt{Frayer:2003p10289, Frayer:2004fk,Clements:2004qy, Pope:2005uq, Hainline:2009kx,Wang:2012ys}). One of the goals of studying the IR properties of the SMGs is to seek color criteria that could efficiently select the correct counterparts to the 850\,$\mu$m SMGs, especially given their large positional uncertainties.

While the NIR color spans a wide range, and there is not yet a single selection method that is successful in choosing the SMGs, a MIR color selection proposed by \citet{Yun:2008vn} using the 8.0\,$\mu$m/4.5\,$\mu$m versus 5.8\,$\mu$m/ 3.6\,$\mu$m flux ratio diagram seems to be able to separate SMGs from other field galaxies. It has been argued that this MIR color selection is too general in that it includes more than 50\% of the field galaxies in the GOODS-N field (\citealt{Hainline:2009kx}). However, it is still a good first order test on our candidate counterparts. We plot any available candidate counterparts from Table~\ref{cntrprt} that have all four {\it Spitzer\/} passband measurements in Figure~\ref{ir_cc}(a). The enclosed area denoted by the solid lines represents the color space of the MIR color selection. We plot the normal field galaxies selected from a deep $K_s$ survey of the GOODS-N field as small gray dots (\citealt{Wang:2010rt}). For visual reference, the MIR measurements on the SMGs observed with the SMA are plotted as triangles for the GOODS-N sample (Barger et al.\ 2012) and asterisks for the AzTEC/COSMOS sample (\citealt{Younger:2007p6982, Younger:2009p9502}). Most of our counterpart candidates, except A370-450.7a and 450.18a, are located within the SMG color regions; thus, A370-450.7a and A370-450.18a could be false counterparts based on this selection.

Another interesting way of selecting SMGs in IR color space is to use the $K_s$--4.5\,$\mu$m color (KIEROs; \citealt{Wang:2010rt, Wang:2012ys}). As opposed to the sources selected by the traditional red color criteria, such as DRGs (Distant Red Galaxies; \citealt{Franx:2003fr}) and EROs (Extremely Red Objects; \citealt{Elston:1988zr, Hu:1994yq}), KIEROs with $K_s-4.5$\,$\mu$m (AB) $>$ 1.6 are mostly dusty starbursting galaxies, and one third of the SMGs that have submillimeter interferometric follow-up are KIEROs (\citealt{Wang:2012ys}). We examine our counterpart candidates on a  $K_s$--4.5\,$\mu$m versus 3.6\,$\mu$m--4.5\,$\mu$m diagram in Figure \ref{ir_cc}(b). This is similar to Figure~1 of Wang et al.\ (2012). We again plot the normal field galaxies selected from a deep $K_s$ survey of the GOODS-N field as small gray dots as in Figure \ref{ir_cc}(a). The SMGs in the GOODS-N field with robust counterpart identifications from Barger et al.\ (2012) are plotted as triangles, and our candidate counterparts are plotted as filled circles. Most of the SMGs cluster at the top-right corner, indicating a very red nature.
Three of our candidate counterparts, including A370-450.7a and A370-450.18a, are located below the SMG locus in $K_s$--4.5\,$\mu$m color, which could mean that either they are simply not the true SMG counterparts or that they are located at lower redshifts (e.g., A370-450.16a is at $z = 0.0286$). While the $K_s$--4.5\,$\mu$m color is an indicator for redshifts (Wang et al.\ 2012), the fact that most of our candidates are located below the KIEROs color cut is evidence that sources detected at both 450\,$\mu$m and 850\,$\mu$m are likely at lower redshifts than sources selected purely at 850\,$\mu$m, which is consistent with our previous results.
 
In Figure \ref{nir_k82}(a), we plot $K_s$ versus the 850\,$\mu$m/20~cm flux ratio using sources having unambiguous radio counterparts in our SCUBA-2 robust sample (filled circles) along with the secure GOODS-N SMG sample with SMA follow-up from Barger et al.\ (2012) (triangles). We find a correlation between these two parameters using the non-parametric Spearman's rank correlation test ($\rho = 0.44$, 15 degree of freedom, P-value$=0.04$), where the higher the submillimeter to radio flux fraction, the fainter the $K_s$ magnitude. We derive a minimum $\chi^2$ fit (black curve) of $log_{10}(f_{850~\mu {\rm m}}/f_{20~cm}) = 0.23\times K_s$(AB) -- 3.24. We also plot $K$-band magnitude versus redshift in Figure \ref{nir_k82}(b) using any spectroscopic redshifts available for the sources in Figure \ref{nir_k82}(a)(including the GOODS-N sample) and millimetric redshifts (shown with error bars) for the ones without spectroscopic redshifts. While the majority of the sources have spectroscopic redshifts, the data distribution is very similar in both panels, indicating that the 850\,$\mu$m/20~cm flux ratio is a good indicator of redshift. The correlation we found could mean that the bright SMG population has a similar rest-frame optical luminosity, and the observed $K_s$-band dimming is caused by the source distance. A similar correlation with $K$-band magnitude versus redshift for SCUBA sources was found by Clements et al.\ (2003) and Pope et al.\ (2005). Our correlation reasonably agrees with their results.

\section{Summary}

In this paper, we present results from our SCUBA-2 450\,$\mu$m and 850\,$\mu$m observations of the massive lensing cluster field A370. Our observations reach a 1\,$\sigma$ sensitivity of 3.92 mJy/beam and 0.82 mJy/beam at 450\,$\mu$m and 850\,$\mu$m, respectively. We summarize our findings below:

1. We detected in total 20 sources at 450\,$\mu$m and 26 sources at 850\,$\mu$m with a significance of $>4\,\sigma$. Using the latest lensing model for A370, we derived the intrinsic submillimeter fluxes and observed number counts. We ran Monte Carlo simulations to obtain the true number counts that take into account lensing effects, completeness, noise distributions, and flux boosting. 

2. We found that our observations resolve $\sim$47--61\% of the 450\,$\mu$m EBL above 4.5~mJy and $\sim$40--57\% of the 850\,$\mu$m EBL above 1.1~mJy. The 450\,$\mu$m counts are the deepest determination to date at this wavelength.

3. We used a statistical method to examine the origin of the 450\,$\mu$m and 850\,$\mu$m emission by plotting the observational positional offsets between the 450\,$\mu$m and 850\,$\mu$m emission against the ideal distribution of offsets, assuming common origins, using our Monte Carlo simulations. We found statistically that the 450\,$\mu$m emission is a good positional tracer for the 850\,$\mu$m population. 

4. We estimated the redshift distributions for our submillimeter samples using millimetric flux ratios assuming a modified blackbody SED template from the local starbursting galaxy Arp 220. We found a median redshift of $\sim2.5$ for only the sources detected at both 450\,$\mu$m and 850\,$\mu$m, $z\sim2.3$ for all the 450\,$\mu$m sources, and $z\sim3.4$ for all the 850\,$\mu$m sources. 

5. The percentage of SMGs having secure radio counterparts is 85\% for 450\,$\mu$m sources with intrinsic fluxes $>6$~mJy and $\sim$67\% for 850\,$\mu$m sources with intrinsic fluxes $>4$~mJy. 67\% of the $>4~\sigma$ 450\,$\mu$m sources are detected in 850\,$\mu$m, while the recovery rate for 850\,$\mu$m sources at 450\,$\mu$m is $\sim$54\%. The recovered rate we found is consistent with the scenario that both 450\,$\mu$m and 20~cm emission trace lower redshift dusty starbursting galaxies, while 850\,$\mu$m emission comes from dusty sources located at higher redshifts.

6. We identified potential counterparts in various wavelengths from the X-ray to the MIR for our SCUBA-2 robust sample (S/N $>$ 5). Three X-ray counterparts are found to be correlated with our robust submillimeter sample. The AGN fraction of our 450\,$\mu$m sample is 3/8, 38\%, and it is 3/9 (33\%) for the 850\,$\mu$m sample. The IR colors of our candidate counterparts mostly agree with the locations of SMGs in previous diagrams, except for a few sources that may be incorrectly identified. We found a correlation between the 850\,$\mu$m/20~cm flux ratio and the $K_s$-band. 
\\

{\it Acknowledgments}
We gratefully acknowledge support from NSF grant AST 0709356 (C.C.C., L.L.C.), the University of Wisconsin Research Committee with funds granted by the Wisconsin Alumni Research Foundation (A.J.B.), the David and Lucile Packard Foundation (A.J.B.), and the National Science Council of Taiwan grant 99-2112-M-001-012-MY3 (W.-H.W.). C.M.C. is generously supported by a Hubble Fellowship provided by Space Telescope Science Institute, grant HST-HF-51268.01-A. We thank the referee for an insightful review, Tim Jenness, Per Friberg, David Berry, Douglas Scott and Edward Chapin for helpful discussions on data reduction, JAC support astronomers Iain Coulson, Holy Thomas and Antonio Chrysostomou, and JCMT telescope operators Jim Hoge, William Montgomerie and Jan Wouterloot. The James Clerk Maxwell Telescope is operated by the Joint Astronomy Centre on behalf of the Science and Technology Facilities Council of the United Kingdom, the Netherlands Organisation for Scientific Research, and the National Research Council of Canada. Additional funds for the construction of SCUBA-2 were provided by the Canada Foundation for Innovation. We acknowledge the cultural significance that the summit of Mauna Kea has to the indigenous Hawaiian community.

\bibliography{bib}

\appendix
\section{Individual sources}

{\it A370-850.1 (450.1)} -- This is the brightest submillimeter source at both 450\,$\mu$m and 850\,$\mu$m in the A370 region and the first submillimeter source discovered by SCUBA (Smail et al.\ 1997). Follow-up observations with high spatial and spectral resolution have revealed its complex nature. Based on several optical and CO spectral line observations (\citealt{Frayer:1998fj,Ivison:1998p10286, Genzel:2003kx}), one of the two optical counterparts, L1, was first identified as the main submillimeter emitter, possibly due to its AGN and star-forming activity, while the nature of the other optical source, L2, was debated between scattered AGN light from L1 (\citealt{Vernet:2001kx}) or the remnant of a merger (Ivison et al.\ 1998). The most recent studies by \citet{Ivison:2010yq} using the CO J=1$\rightarrow$0 line and data newly obtained from {\it Hubble\/} and {\it Spitzer\/} argue that a heavily dust-obscured starburst between L1 and L2, L2SW, is more likely to be the true counterpart. They also argue that together with a bright UV emitter, L1N, this could be a merger system with at least two galaxies. However, due to the positional uncertainties of all the sources, it is hard to rule out the scenario proposed by Genzel et al.\ (2003) of a gas reservoir residing in a massive, extended disk around L1. Interestingly, our 450\,$\mu$m emission peaks on L1, while the 850\,$\mu$m emission peaks on L2SW. Based on our results, we argue that both L1 and L2SW contribute to the FIR emission. Future subarcsecond observations using the EVLA or ALMA will be the key to fully understanding a system with such complexity. For simplicity, in this paper we only provide the photometric information for L1.

{\it A370-850.2 (450.3)} -- With the help of better spatial resolution data at 450\,$\mu$m, an optically faint, MIR and radio bright counterpart is clearly identified. The redshift estimate for this source using the 850\,$\mu$m-to-20~cm flux ratio (see Section~5.3) is $\sim1.90\pm0.12$.

{\it A370-850.3 (450.2)} -- This source was also detected in a SCUBA survey at both 450\,$\mu$m and 850\,$\mu$m (Smail et al.\ 2002). Two possible counterparts are detected across all bandpasses. Detailed observations with the Keck~II LRIS spectrograph show that L3, an AGN at $z=1.06$ with a ring morphology, is more likely to be the counterpart than L5, a passive cluster elliptical (\citealt{Barger:1999p6801}). This was confirmed by high-resolution interferometric observations using the SMA (Chen et al.\ 2011), and our high S/N 450\,$\mu$m detection confirms this result. 

{\it A370-850.4 (450.17)} -- Both a 5\,$\sigma$ and a 4\,$\sigma$ radio source are located within the submillimeter positional errors. Both sources are optically and IR faint and therefore very dusty. It is likely that both radio sources contribute part of the submillimeter emission.

{\it A370-850.5 (450.9/450.11)} -- The submillimeter morphology of A370-850.5 extends towards A370-450.9. We subtracted the PSF centered at the peak position of A370-850.5 to obtain the 850\,$\mu$m flux for A370-450.9, which is 3.91$\pm$1.49~mJy. A 3\,$\sigma$ radio source sits close to the A370-450.9 position with faint emission at 4.5\,$\mu$m. A bright radio source candidate counterpart (A370-450.11a) with a spectroscopic redshift of $z=3.8219$ is located on the verge of the 95\% confidence positional area. Another candidate counterpart (A370-450.11b) is a faint $z$-band source. This region is the most complex system in our survey with potentially up to three sources (A370-450.9, A370-450.11, and A370-850.5) contributing to the submillimeter emission.

{\it A370-850.6 (450.5)} -- One {\it Spitzer\/} source is located within the submillimeter positional error. It has a 3\,$\sigma$ radio counterpart and is not visible in the optical and NIR. From the 850\,$\mu$m-to-20~cm flux ratio, we estimated a redshift of $z \sim 4.47$ for this source.
 
{\it A370-850.7 (450.18)} -- A bright galaxy candidate counterpart (A370-450.18a) is located very close to the 850\,$\mu$m position. This galaxy was reported in earlier optical and NIR cluster survey work (\citealt{Stanford:1995yq, Margoniner:2000vn}). We estimated the redshift of the SMG to be $z\sim5$ from the 850\,$\mu$m-to-450\,$\mu$m flux ratio. Although there is no redshift measurement for this source, the fact that it is resolved in the optical and IR makes it likely to be at low redshifts. In addition, the fact that the SED of A370-450.18a decays towards longer wavelengths makes it unlikely to be the origin of the submillimeter emission. More tentative candidate counterparts are selected within the 450\,$\mu$m positional error, where there are two $>3.5~\sigma$ optically faint, IR faint, radio detections. If those two radio sources are proven to be the correct counterparts, then they would be great examples of how 450\,$\mu$m positions can be used to select counterparts to 850\,$\mu$m sources.

{\it A370-850.8 (450.7)} -- The candidate counterpart A370-450.7a is seen in the optical and IR and was detected in other optical ($g, r, i$) surveys (Margoniner \& Carvalho  2000). Faint emission in the {\it Spitzer\/} passbands can be seen towards the two faint radio source candidate counterparts, A370-450.7b and A370-450.7c. We estimated a redshift of $z\sim2.5$ for the SMG using the 850\,$\mu$m-to-450\,$\mu$m flux ratio.

{\it A370-850.9 (450.15)} --  One optically faint but MIR and radio bright counterpart is identified. We estimated a redshift of $\sim2.29\pm0.3$ for this source using the 850\,$\mu$m-to-20~cm flux ratio.

{\it A370-850.10 (450.16)} -- A bright local galaxy candidate counterpart (A370-450.16a) with $z=0.0286$ is seen in the north-east. This source was reported in various observations from the optical to the X-ray (Margoniner \& Carvalho  2000; \citealt{Barger:2001uq, Metcalfe:2003lp}). Although its IR color does not put it around the locus of a typical SMG, we cannot rule out the possibility of A370-450.16a contributing to the submillimeter emission, since it is bright in the radio. Meanwhile, another radio source candidate counterpart (A370-450.16b) is located in the south-west. Its optical and IR faint nature makes it another plausible counterpart. It is likely that both A370-450.16a and A370-450.16b contribute part of the submillimeter emission.

{\it A370-850.18 (450.4)} --  An optically faint, MIR and radio bright counterpart is clearly identified by the 450\,$\mu$m emission. This source has a redshift of $z=1.519$ (Wold et al.\ 2012), which agrees  within the errors with our redshift estimations for this source using the flux ratios. 

\end{document}